%% file: main.tex
\documentclass[times,onecolumn]{aastex62}

\usepackage{cases}
\usepackage{graphicx}
\usepackage{subfigure}
\usepackage{natbib}
\usepackage{color}
\usepackage{tabularx}
\usepackage{bm}
\usepackage{threeparttable}

\newcommand{\thetae}{\theta_{\rm E}}
\newcommand{\pie}{\pi_{\rm E}}
\newcommand{\pis}{\pi_{\rm S}}

\newcommand{\te}{t_{\rm E}}
\newcommand{\event}{KMT-2020-BLG-0414}

\newcommand{\qe}{q_{\oplus}}

\DeclareGraphicsExtensions{.pdf,.png,.jpg}

\shorttitle{}
\shortauthors{Zang et al.}

\begin{document}
\title{{\large An Earth-mass Planet in a Time of Covid-19: KMT-2020-BLG-0414Lb}}

\correspondingauthor{Weicheng Zang, Cheongho Han}
\email{zangwc17@mails.tsinghua.edu.cn, cheongho@astroph.chungbuk.ac.kr}

\input{author.tex}

\input{abstract}

\section{Introduction}\label{intro}
\input{intro}

\section{Observations}\label{obser}
\input{obser}

\section{2L1S Analysis}\label{2L1S}

\input{2L1S}

\section{3L1S Analysis}\label{3L1S}

\input{3L1S}

\section{Physical Parameters}\label{lens}

\input{lens}

\section{Discussion}\label{dis}
\input{dis}

\bibliography{Zang.bib}

\input{table.tex}

\input{figure.tex}

\end{document}

%% file: author.tex
\author[0000-0001-6000-3463]{Weicheng Zang}
\affiliation{Department of Astronomy, Tsinghua University, Beijing 100084, China}

\author{Cheongho Han}
\affiliation{Department of Physics, Chungbuk National University, Cheongju 28644, Republic of Korea}

\author{Iona~Kondo}
\affiliation{Department of Earth and Space Science, Graduate School of Science, Osaka University, Toyonaka, Osaka 560-0043, Japan}

\author{Jennifer C. Yee}
\affiliation{Center for Astrophysics $|$ Harvard \& Smithsonian, 60 Garden St.,Cambridge, MA 02138, USA}

\author{Chung-Uk Lee}
\affiliation{Korea Astronomy and Space Science Institute, Daejon 34055, Republic of Korea}
\affiliation{University of Science and Technology, Korea, (UST), 217 Gajeong-ro Yuseong-gu, Daejeon 34113, Republic of Korea}

\author{Andrew Gould}
\affiliation{Max-Planck-Institute for Astronomy, K\"onigstuhl 17, 69117 Heidelberg, Germany}
\affiliation{Department of Astronomy, Ohio State University, 140 W. 18th Ave., Columbus, OH 43210, USA}

\author{Shude Mao}
\affiliation{Department of Astronomy, Tsinghua University, Beijing 100084, China}
\affiliation{National Astronomical Observatories, Chinese Academy of Sciences, Beijing 100101, China}

\author{Leandro de Almeida}
\affiliation{Universidade Federal do Rio Grande do Norte (UFRN), Departamento de F\'1sica, 59078-970, Natal, RN, Brazil}

\author{Yossi Shvartzvald}
\affiliation{Department of Particle Physics and Astrophysics, Weizmann Institute of Science, Rehovot 76100, Israel}

\author{Xiangyu Zhang}
\affiliation{Department of Astronomy, Tsinghua University, Beijing 100084, China}

\collaboration{(Leading Authors)}


\author{Michael D. Albrow}
\affiliation{University of Canterbury, Department of Physics and Astronomy, Private Bag 4800, Christchurch 8020, New Zealand}

\author{Sun-Ju Chung}
\affiliation{Korea Astronomy and Space Science Institute, Daejon 34055, Republic of Korea}
\affiliation{University of Science and Technology, Korea, (UST), 217 Gajeong-ro Yuseong-gu, Daejeon 34113, Republic of Korea}

\author{Kyu-Ha Hwang}
\affiliation{Korea Astronomy and Space Science Institute, Daejon 34055, Republic of Korea}

\author{Youn Kil Jung}
\affiliation{Korea Astronomy and Space Science Institute, Daejon 34055, Republic of Korea}

\author{Yoon-Hyun Ryu}
\affiliation{Korea Astronomy and Space Science Institute, Daejon 34055, Republic of Korea}

\author{In-Gu Shin}
\affiliation{Korea Astronomy and Space Science Institute, Daejon 34055, Republic of Korea}

\author{Sang-Mok Cha}
\affiliation{Korea Astronomy and Space Science Institute, Daejon 34055, Republic of Korea}
\affiliation{School of Space Research, Kyung Hee University, Yongin, Kyeonggi 17104, Republic of Korea} 

\author{Dong-Jin Kim}
\affiliation{Korea Astronomy and Space Science Institute, Daejon 34055, Republic of Korea}

\author{Hyoun-Woo Kim}
\affiliation{Korea Astronomy and Space Science Institute, Daejon 34055, Republic of Korea}

\author{Seung-Lee Kim}
\affiliation{Korea Astronomy and Space Science Institute, Daejon 34055, Republic of Korea}
\affiliation{University of Science and Technology, Korea, (UST), 217 Gajeong-ro Yuseong-gu, Daejeon 34113, Republic of Korea}

\author{Dong-Joo Lee}
\affiliation{Korea Astronomy and Space Science Institute, Daejon 34055, Republic of Korea}

\author{Yongseok Lee}
\affiliation{Korea Astronomy and Space Science Institute, Daejon 34055, Republic of Korea}
\affiliation{School of Space Research, Kyung Hee University, Yongin, Kyeonggi 17104, Republic of Korea}

\author{Byeong-Gon Park}
\affiliation{Korea Astronomy and Space Science Institute, Daejon 34055, Republic of Korea}
\affiliation{University of Science and Technology, Korea, (UST), 217 Gajeong-ro Yuseong-gu, Daejeon 34113, Republic of Korea}

\author{Richard W. Pogge}
\affiliation{Department of Astronomy, Ohio State University, 140 W. 18th Ave., Columbus, OH  43210, USA}

\collaboration{(The KMTNet Collaboration)}

\author{John Drummond}
\affiliation{Possum Observatory, Patutahi, Gisbourne, New Zealand}

\author{Thiam-Guan Tan}
\affiliation{Perth Exoplanet Survey Telescope, Perth, Australia}

\author{Jos\'e Dias do Nascimento J\'unior}
\affiliation{Universidade Federal do Rio Grande do Norte (UFRN), Departamento de F\'1sica, 59078-970, Natal, RN, Brazil}
\affiliation{Center for Astrophysics $|$ Harvard \& Smithsonian, 60 Garden St.,Cambridge, MA 02138, USA}

\author{Dan Maoz}
\affiliation{School of Physics and Astronomy, Tel-Aviv University, Tel-Aviv 6997801, Israel}

\author[0000-0001-7506-5640]{Matthew T. Penny}
\affiliation{Department of Physics and Astronomy, Louisiana State University, Baton Rouge, LA 70803 USA}

\author{Wei Zhu}
\affiliation{Department of Astronomy, Tsinghua University, Beijing 100084, China}
\affiliation{Canadian Institute for Theoretical Astrophysics, University of Toronto, 60 St George Street, Toronto, ON M5S 3H8, Canada}

\collaboration{(The LCO \& $\mu$FUN Follow-up Teams)}

\author{Ian~A.~Bond}
\affiliation{Institute of Natural and Mathematical Sciences, Massey University, Auckland 0745, New Zealand}

\author{Fumio~Abe}
\affiliation{Institute for Space-Earth Environmental Research, Nagoya University, Nagoya 464-8601, Japan}

\author{Richard Barry}
\affiliation{Code 667, NASA Goddard Space Flight Center, Greenbelt, MD 20771, USA}

\author{David~P.~Bennett}
\affiliation{Code 667, NASA Goddard Space Flight Center, Greenbelt, MD 20771, USA}
\affiliation{Department of Astronomy, University of Maryland, College Park, MD 20742, USA}

\author{Aparna~Bhattacharya}
\affiliation{Code 667, NASA Goddard Space Flight Center, Greenbelt, MD 20771, USA}
\affiliation{Department of Astronomy, University of Maryland, College Park, MD 20742, USA}

\author{Martin~Donachie}
\affiliation{Department of Physics, University of Auckland, Private Bag 92019, Auckland, New Zealand}

\author{Hirosane Fujii}
\affiliation{Department of Earth and Space Science, Graduate School of Science, Osaka University, Toyonaka, Osaka 560-0043, Japan}

\author{Akihiko~Fukui}
\affiliation{Department of Earth and Planetary Science, Graduate School of Science, The University of Tokyo, 7-3-1 Hongo, Bunkyo-ku, Tokyo 113-0033, Japan}
\affiliation{Instituto de Astrof\'isica de Canarias, V\'ia L\'actea s/n, E-38205 La Laguna, Tenerife, Spain}

\author{Yuki~Hirao}
\affiliation{Department of Earth and Space Science, Graduate School of Science, Osaka University, Toyonaka, Osaka 560-0043, Japan}

\author{Yoshitaka~Itow}
\affiliation{Institute for Space-Earth Environmental Research, Nagoya University, Nagoya 464-8601, Japan}

\author{Rintaro Kirikawa}
\affiliation{Department of Earth and Space Science, Graduate School of Science, Osaka University, Toyonaka, Osaka 560-0043, Japan}

\author{Naoki~Koshimoto}
\affiliation{Department of Astronomy, Graduate School of Science, The University of Tokyo, 7-3-1 Hongo, Bunkyo-ku, Tokyo 113-0033, Japan}
\affiliation{National Astronomical Observatory of Japan, 2-21-1 Osawa, Mitaka, Tokyo 181-8588, Japan}

\author{Man~Cheung~Alex~Li}
\affiliation{Department of Physics, University of Auckland, Private Bag 92019, Auckland, New Zealand}

\author{Yutaka~Matsubara}
\affiliation{Institute for Space-Earth Environmental Research, Nagoya University, Nagoya 464-8601, Japan}

\author{Yasushi~Muraki}
\affiliation{Institute for Space-Earth Environmental Research, Nagoya University, Nagoya 464-8601, Japan}

\author{Shota~Miyazaki}
\affiliation{Department of Earth and Space Science, Graduate School of Science, Osaka University, Toyonaka, Osaka 560-0043, Japan}

\author{Cl\'ement~Ranc}
\affiliation{Code 667, NASA Goddard Space Flight Center, Greenbelt, MD 20771, USA}

\author{Nicholas~J.~Rattenbury}
\affiliation{Department of Physics, University of Auckland, Private Bag 92019, Auckland, New Zealand}

\author{Yuki Satoh}
\affiliation{Department of Earth and Space Science, Graduate School of Science, Osaka University, Toyonaka, Osaka 560-0043, Japan}

\author{Hikaru Shoji}
\affiliation{Department of Earth and Space Science, Graduate School of Science, Osaka University, Toyonaka, Osaka 560-0043, Japan}

\author{Takahiro Sumi}
\affiliation{Department of Earth and Space Science, Graduate School of Science, Osaka University, Toyonaka, Osaka 560-0043, Japan}

\author{Daisuke~Suzuki}
\affiliation{Institute of Space and Astronautical Science, Japan Aerospace Exploration Agency, 3-1-1 Yoshinodai, Chuo, Sagamihara, Kanagawa, 252-5210, Japan}

\author{Yuzuru Tanaka}
\affiliation{Department of Earth and Space Science, Graduate School of Science, Osaka University, Toyonaka, Osaka 560-0043, Japan}

\author{Paul~J.~Tristram}
\affiliation{University of Canterbury Mt.\ John Observatory, P.O. Box 56, Lake Tekapo 8770, New Zealand}

\author{Tsubasa Yamawaki}
\affiliation{Department of Earth and Space Science, Graduate School of Science, Osaka University, Toyonaka, Osaka 560-0043, Japan}

\author{Atsunori~Yonehara}
\affiliation{Department of Physics, Faculty of Science, Kyoto Sangyo University, 603-8555 Kyoto, Japan}
\collaboration{(The MOA Collaboration)}

\author{Andreea Petric}
\affiliation{CFHT Corporation, 65-1238 Mamalahoa Hwy, Kamuela, Hawaii 96743, USA}
\affiliation{Space Telescope Science Institute, Baltimore, MD 21211}

\author{Todd Burdullis}
\affiliation{CFHT Corporation, 65-1238 Mamalahoa Hwy, Kamuela, Hawaii 96743, USA}

\author{Pascal Fouqu\'e}
\affiliation{CFHT Corporation, 65-1238 Mamalahoa Hwy, Kamuela, Hawaii 96743, USA}
\affiliation{Universit\'e de Toulouse, UPS-OMP, IRAP, Toulouse, France}

\collaboration{(CFHT Microlensing Collaboration)}

%% file: abstract.tex
\begin{abstract}
We report the discovery of KMT-2020-BLG-0414Lb, with a planet-to-host mass ratio $q_2 = 0.9$--$1.2 \times 10^{-5} = 3$--$4~\qe$ at $1\sigma$, which is the lowest mass-ratio microlensing planet to date. Together with two other recent discoveries ($4 \lesssim q/q_\oplus \lesssim 6$), it fills out the previous empty sector at the bottom of the triangular $(\log s, \log q)$ diagram, where $s$ is the planet-host separation in units of the angular Einstein radius $\thetae$. Hence, these discoveries call into question the existence, or at least the strength, of the break in the mass-ratio function that was previously suggested to account for the paucity of very low-$q$ planets. Due to the extreme magnification of the event, $A_{\rm max}\sim 1450$ for the underlying single-lens event, its light curve revealed a second companion with $q_3 \sim 0.05$ and $|\log s_3| \sim 1$, i.e., a factor $\sim 10$ closer to or farther from the host in projection. The measurements of the microlens parallax $\bm{\pi}_{\rm E}$ and the angular Einstein radius $\thetae$ allow estimates of the host, planet, and second companion masses, $(M_1, M_2, M_3) \sim (0.3M_{\odot}, 1.0M_{\oplus}, 17M_{J})$, the planet and second companion projected separations, $(a_{\perp,2}, a_{\perp,3}) \sim (1.5, 0.15~{\rm or}~15)$~au, and system distance $D_{\rm L} \sim 1$ kpc. The lens could account for most or all of the blended light ($I \sim 19.3$) and so can be studied immediately with high-resolution photometric and spectroscopic observations that can further clarify the nature of the system. The planet was found as part of a new program of high-cadence follow-up observations of high-magnification events. The detection of this planet, despite the considerable difficulties imposed by Covid-19 (two KMT sites and OGLE were shut down), illustrates the potential utility of this program.
\end{abstract}


%% file: intro.tex
It has long been known that intensive monitoring of high-magnification microlensing events is sensitive to planets of one-to-few Earth/Sun
mass ratio, $\qe = 3 \times 10^{-6}$, planets. \cite{OB04343} showed that the $A_{\rm max} \sim 3000$ event OGLE-2004-BLG-343 would have had such sensitivity had it been observed over peak (see their Figure 9). \cite{OB08279} showed such sensitivity for the actual data covering the peak of the $A_{\rm max} \sim 1600$ event OGLE-2008-BLG-279 (see their Figure 7). In both cases (and, indeed, for high magnification events in general, \citealt{Gaudi2002,mufun}), the sensitivity diagrams have a triangular appearance that is symmetric in $\log s$ about the origin.  That is, the contour limits meet at $s = 1$, where $s$ is the planet-host separation in units of the Einstein radius $\thetae$. Hence, the limiting sensitivity in $q$ is via a so-called ``resonant caustic''.  For $s \gg 1$, the caustic structure consists of a small quadrilateral caustic near the host and a larger quadrilateral caustic near the planet.  For $s \ll 1$, it consists of a similar small quadrilateral caustic near the host and two triangular caustics located on the opposite side of the planet.  For $s \sim 1$, these two sets of caustics merge into a relatively large 6-sided ``resonant'' caustic, which is what makes the detection feasible at the very limit of sensitivity\footnote{In fact, \citet{OB190960} showed that ``semi-resonant'' caustics, which have not quite merged but still have long magnification ridges extending from the central caustic, as well as exceptionally large planetary caustics, are just as sensitive as resonant caustics.}.

Neither of the above two events yielded any planet detection, but \cite{OB171434} did find a resonant-caustic planet in OGLE-2017-BLG-1434 ($A_{\rm max} \sim 23$ for the underlying single-lens event). While its mass ratio was $q \sim 5.7 \times 10^{-5}$ (i.e., $q \sim 19~\qe$), \cite{OB171434} showed that a planet with exactly the same characteristics, but 30 times less massive, would have been detected (see their Figure 4).

Nevertheless, in spite of the recognized theoretical possibility of such few-$\qe$ planet detections, no planet with mass ratio below that of Uranus/Sun
mass ratio $q \sim 15~\qe$ was actually detected prior to 2018\footnote{Although OGLE-2017-BLG-0173 has a best-fit solution of $q \sim 8~\qe$, its two degenerate solutions have  $q \sim 21~\qe$ at $\Delta\chi^2 = 3.5$ \citep{OB170173}.}. This failure gave rise to several different suggestions of a paucity of low mass-ratio planets. \cite{Suzuki2016} argued for a ``break'' in the mass-ratio function at $q_{\rm br} \sim 57~\qe$ based on a statistically well-defined sample of planets detected from MOA-discovered microlensing events. \cite{KB170165} argued for a ``break'', or possibly a ``pile-up'' at $q \sim 19~\qe$ based on the ensemble of planets with $q< 3 \times 10^{-4}$. \cite{OB171434} used a new ``$V/V_{\rm max}$'' method to show that if planets with $q < 10^{-4}$ were modeled as a power-law distribution in $q$, then the distribution was rising toward higher $q$, seemingly confirming the \cite{Suzuki2016} ``break''. 

However, in 2018 and 2019, three planets were discovered with $q$ below the previous record, and hence below the level of the conjectured ``break'' or ``pile-up'': KMT-2018-BLG-0029Lb ($q \sim 1.8 \times 10^{-5}$, \citealt{KB180029}), KMT-2019-BLG-0842Lb ($q \sim 4.1 \times 10^{-5}$, \citealt{KB190842}), and OGLE-2019-BLG-0960Lb ($q \sim 1.4 \times 10^{-5}$, \citealt{OB190960}). All three were detected via resonant caustics at or near the peak of moderately high magnification events, with $|s-1|\sim(0.000,0.017,0.003~{\rm or}~0.029)$. Moreover, \citet{OB190960} showed that the recent discoveries of KMT-2018-BLG-0029Lb and OGLE-2019-BLG-0960Lb populate the previously-vacant lower region of the $(\log s,\log q)$ diagram. See their Figure 11. It is therefore clear that the well-established sensitivity to $q \sim \qe$ depends primarily on relatively rare (i.e., high-magnification) microlensing events generated by relatively rare resonant lens configurations. And, therefore, it is possible that the previous paucity of planets near the detection limit was more a product of the difficulty of detection than the intrinsic rarity of the population.

In this context, it is notable that over the last 10 years, microlensing planet searches have moved away from their previous focus on high-magnification events, which is one of the two ``rare elements'' just described for probing the low-$q$ population. Prior to the inauguration of the wide-field-camera OGLE-IV
survey \citep{OGLEIV}, substantial effort was expended, particularly by the Microlensing Follow Up Network ($\mu$FUN), to find high-magnification events, and then to focus intensive observations over the peak \citep{mufun}. As a result, the projected spatial distribution of planetary events was drawn roughly uniformly from the OGLE-III and MOA-II surveys. See the blue circles in Figure 8 of \cite{KB181292}. That is, the surveys were able to detect events, more or less regardless of cadence, but were much less able to detect planets on their own, again regardless of cadence. Hence, planet-yielding events were unaffected by survey cadence. However, with the layered approach (higher cadence in more productive fields) made possible by the introduction of the larger-format OGLE-IV camera, planet searches came to rely more on survey data, so that planetary discoveries became more concentrated on high-cadence regions. See green and yellow points of the same diagram. This remained so with the advent of the still larger-format KMTNet \citep{KMT2016} survey (magenta and black points).

Beginning in 2016, KMTNet continuously monitors $\sim 97~{\rm deg}^2$ area of sky toward the Galactic bulge field from three 1.6m telescopes equipped with 4 ${\rm deg}^2$ FOV cameras at CTIO in Chile, SAAO in South Africa, and SSO in Australia. In fact, KMTNet's three-observatory system is capable of detecting very low-$q$ planets in its highest cadence, $\Gamma = 4\,{\rm hr}^{-1}$ fields, as was proved by the case of KMT-2019-BLG-0842Lb. The substantially lower-$q$ planet KMT-2018-BLG-0029Lb was discovered in KMT-only observations of a $\Gamma = 1\,{\rm hr}^{-1}$ field. For the lowest-$q$ planet OGLE-2019-BLG-0960Lb, although the planetary signal was first recognized by the $\mu$FUN CT13 data and was most clearly delineated by the $\mu$FUN Kumeu data, the detection would probably have been regarded as secure using the low-cadence survey observations ($\Gamma_{\rm KMT} \sim 0.4\,{\rm hr}^{-1}$, 3 sites; $\Gamma_{\rm OGLE} = 0.17\,{\rm hr}^{-1}$, 1 site; $\Gamma_{\rm MOA} = 0.6\,{\rm hr}^{-1}$, 1 site).

An additional notable feature of these detections is that the magnification of the underlying 1L1S event at the time of the planetary anomaly was modest ($A_{\rm anom} \sim 37$ for KMT-2018-BLG-0029Lb, $A_{\rm anom} \sim 22$ for KMT-2019-BLG-0842Lb and $A_{\rm anom} \sim 45$ for OGLE-2019-BLG-0960Lb),  although $A_{\rm max} \sim 160$ for the latter two cases. The source trajectory was ``oblique'', i.e., close to parallel to the long axis of the resonant caustic, with $\alpha = 8.3^\circ$ and $\alpha = 15.5^\circ$ for KMT-2019-BLG-0842Lb and OGLE-2019-BLG-0960Lb, respectively. While such oblique trajectories are relatively rare, they can enhance detection efficiency by ``stretching out'' anomalies. Moreover, \cite{OB190960} found that sensitivity to very low-$q$ planets can be maximized by intensively monitoring events whenever they were magnified by a factor $A > 10$ (see also \citealt{Abe2013}). In order to probe the very low-$q$ planets, KMTNet together with the LCO \& $\mu$FUN Follow-up Team developed a program for focusing on observations and analysis of $A \gtrsim 20$ events located in KMTNet $\Gamma \leq 1\,{\rm hr}^{-1}$ fields. 

However, with the advent of Covid-19, two of KMT's three observatories were shut down, leaving only KMT's Australia telescope as operational. Hence, the conditions became much more similar to those of the first decade of this century, when microlensing alerts came primarily from the OGLE-III survey, and planets were primarily discovered by follow-up observations of these single-site alerts, as well as some non-overlapping alerts from MOA.  Of course, there were some differences.  In particular, KMT has a much larger format camera than OGLE-III, and so it operates at substantially higher cadence. However, KMTA also has the worst conditions of KMT's three observatories, and so is inferior to OGLE-III in both weather interruptions and photometric precision. 

It was the specific orientation of this program, i.e., follow-up observations of $A \gtrsim 20$ events that led to the discovery of KMT-2020-BLG-0414Lb, the lowest mass-ratio microlensing planet discovered to date, $q_2 \sim 10^{-5}$. Due to the extreme magnification of the event, $A_{\rm max} \sim 1450$ for the underlying single-lens event, and high-cadence observations by MOA over the peak, a second companion was also detected, with $q_3 \sim 0.05$.

%% file: obser.tex
\event\ occurred at equatorial coordinates $(\alpha, \delta)_{\rm J2000}$ = (18:07:39.60, $-$28:29:06.8), corresponding to Galactic coordinates $(\ell,b)=(2.82, -3.95)$. It was announced as a ``probable microlensing'' event by the KMTNet Alert-Finder system \citep{KMTAF} on 1 June 2020, about 40 days before peak, when the event was manifested as an $I\sim 18.7$ difference star. As mentioned in Section \ref{intro}, by this date, KMTC and KMTS had been closed down for more than two months due to Covid-19.  Hence, only KMTA data contributed to the alert and to subsequent monitoring of the event. The event lies in the KMNTet BLG32 field, which has a cadence of $\Gamma = 0.4~{\rm hr}^{-1}$, with every tenth $I$-band observation complemented by one in the $V$ band for the source color measurements\footnote{In fact, this $V$-band to $I$-band ratio applies only to the ``normal cycle'' of KMT observations. During the latter part of the season (including the peak of KMT-2020-BLG-0414), these normal-cycle observations were supplemented by an end-of-night sequence of Eastern fields, which was purely in the $I$-band. For the low-cadence field BLG32, these end-of-night observations accounted for $\sim 30\%$ of the total near the peak of the event.}. 

The event was independently identified by the Microlensing Observations in Astrophysics (MOA, \citealt{MOA2016}) collaboration as MOA-2020-BLG-109 on 20 June 2020 \citep{Bond2001}. MOA observations are carried out with a 1.8m telescope at the Mt.\ John University Observatory in New Zealand, which is equipped with a 2.2 square degree camera.  The nominal cadence for this field was $\Gamma = 1.2~{\rm hr}^{-1}$ by a MOA-Red filter (which is similar to the sum of the standard Cousins $R$- and $I$-band filters), and observations with the MOA $V$ filter (Bessell V-band) were taken once every clear night. Earlier in the season, the MOA survey had also been closed for Covid-19 for almost 100 days, but it had re-opened 28 days before peak. Hence, the Covid-19 hiatus had very little effect on its observations of this event.

The KMTNet Alert-Finder system identified \event\ with a catalog star at $(\alpha, \delta)_{\rm J2000}$ = (18:07:39.60,$-28$:29:05.50), which is about $1.3^{\prime\prime}$ away from the true position of the source. As a result, the real-time on-line photometry was relatively noisy. Nevertheless, at UT 17:34 on 7 July 2020 (${\rm HJD}^{\prime} = 9038.23, {\rm HJD}^{\prime} = {\rm HJD} - 2450000$), the LCO \& $\mu$FUN Follow-up Team found that this event had magnification $A_{\rm now} > 10$ based on the two KMTA points at ${\rm HJD}^{\prime} \sim 9038$ and could peak at a high magnification 2--3 days later\footnote{The LCO \& $\mu$FUN Follow-up Team recognized the KMTA photometric centroid shift by its noisy curve and thus started follow-up observations for security, although $A_{\rm now}$ did not meet the $A \gtrsim 20$ threshold. In fact, the actual magnification at that time was about 35.}. Thus, high-cadence follow-up observations were immediately scheduled by Las Cumbres Observatory (LCO) global network and Observatorio do Pico dos Dias (OPD) in Brazil (a $\mu$FUN site). The LCO global network conducted observations from its 1.0m telescopes located at SAAO (LCOS), SSO (LCOA) and McDonald (LCOM), with the SDSS-$i'$ filter. Observations by OPD were taken from its 0.6m (OPD06) and 1.6m (OPD16) telescopes with the $I$ filter. At UT 05:58 on 10 July 2020 (${\rm HJD}^{\prime} = 9040.75$), the LCO \& $\mu$FUN Follow-up Team identified that this event was currently undergoing an anomaly and would peak at a very high magnification soon, based on the real-time LCO and MOA data. Noting that SSO was predicted to be rainy that night, the Team issued an alert to the MOA collaboration. MOA responded to the alert and densely observed this event 60 times over the peak. Due to the very high brightness, the MOA observer decreased the exposure time from 60s to 5.2s over the peak. We carefully inspected these 60 MOA images and excluded 19 data points from the analysis due to saturation or bad seeing. At UT 13:53 on 10 July 2020 (${\rm HJD}^{\prime} = 9041.08$), the LCO \& $\mu$FUN Follow-up Team also issued an alert to all $\mu$FUN observers. As a result, the 0.3m Perth Exoplanet Survey Telescope (PEST) in Australia and the 0.4m Possum Observatory (Possum) at New Zealand responded to the alert and took intensive observations without a filter. Finally, the event was also observed by the 3.6m Canada-France-Hawaii Telescope (CFHT) with the SDSS-$i'$ filter. 

For the light curve analysis, the KMTNet, MOA, CFHT and LCO data were reduced using custom implementations of the difference image analysis technique \citep{Tomaney1996,Alard1998}: pySIS \citep{pysis} for the KMTNet data, \cite{Bond2001} for the MOA data and ISIS \citep{Alard1998,Alard2000,CFHT} for the CFHT and LCO data. The OPD, PEST and Possum data were reduced using \texttt{DoPHOT}~\citep{dophot}. On the OPD16 images, the target was affected by a bleed trail from a saturated star, resulting in some systematics. We therefore do not include OPD16 data in the analysis. The $I$-band magnitude of the KMTA light curve has been calibrated to the standard $I$-band magnitude. For the source color measurements, we use the MOA $V$-band data, while KMTA $V$-band data are not used due to poor seeing. The errors from photometric measurements for each data set $i$ were renormalized using the formula $\sigma_i^{\prime} = k_i\sqrt{\sigma_i^2 + e^2_{i,{\rm min}}}$, where $\sigma_i$ and $\sigma_i^{\prime}$ are original errors from the photometry pipelines and renormalized error bars in magnitudes, and $k_i$ and $e_{i,{\rm min}}$ are rescaling factors. We obtained the rescaling factors using the procedure of \cite{MB11293}, which enables $\chi^2/{\rm dof}$ for each data set to become unity. We derived the rescaling factors using the binary lens (2L1S) and triple lens (3L1S) models, respectively, in order to understand how the event would have been interpreted in the absence of MOA data on the peak. The data used in the analysis, together with corresponding data reduction method and rescaling factors are summarized in Table \ref{data}. 


%% file: 2L1S.tex
Figure \ref{lc1} shows the observed light curve of \event. Although the light curve can be regarded as ``single peak'' in the sense that it monotonically rises and then falls, its significantly asymmetric shape cannot be fitted by a single-lens single-source (1L1S) model. A 1L1S model is usually described by three \citet{Paczynski1986} parameters $(t_0, u_0, \te)$, i.e., the time of closest lens-source approach, the impact parameter scaled to $\thetae$, and the Einstein crossing time,
\begin{equation}\label{eqn:1}
\te = \frac{\thetae}{|\bm{\mu}_{\rm rel}|}; \qquad \thetae = \sqrt{\kappa M_{\rm L} \pi_{\rm rel}}; \qquad \kappa \equiv \frac{4G}{c^2\mathrm{au}} \simeq 8.144 \frac{{\rm mas}}{M_{\odot}},
\end{equation}  
where $M_{\rm L}$ is the mass of the lens and $(\pi_{\rm rel},\bm{\mu}_{\rm rel})$ are the lens-source relative (parallax, proper motion).  In the present
case, we also consider finite-source effects \citep{1994ApJ...421L..75G,Shude1994,Nemiroff1994}, which occur when the source passes close to singular structures in the magnification pattern. This requires a fourth parameter $\rho = \theta_* / \thetae$, where $\theta_*$ is the angular radius of the source.

As we will show in Section \ref{3L1S}, the full light curve cannot be explained by a 2L1S model, and in fact requires 3L1S. However, if we exclude the MOA data on the peak ($9040.78 < {\rm HJD}^{\prime} < 9041.20$, the only data set to cover the peak), then the remaining data are quite well fit by a 2L1S model. We therefore begin by analyzing this restricted (non-MOA-peak) data set. We are motivated by two considerations. First, and most importantly, 3L1S models often ``factor'' into two 2L1S models \citep{Han2005,OB06109,OB130341,OB120026,OB160613,OB181011}. In particular, the 3L1S caustic is often very nearly the superposition of the two 2L1S caustics.  In such cases, one can often exclude the data from the neighborhood of the anomaly from the ``second'' 2L1S model in order to accurately determine the parameters of the ``first'' 2L1S model. In these cases, the ``first'' 2L1S model provides a powerful basis for finding the full 3L1S model, by one of several techniques. This proves to be the case for \event. Second, it is of independent scientific interest to understand how the event would have been interpreted and check the so-called ``higher-order effects'' in the absence of MOA data on the peak. For example, MOA could have been weathered out on the night of the peak (as was KMTA). The analysis would have led to a report of a single low-mass-ratio planet and a lens that is much brighter than the blended light. The comparison of this reconstructed ``report'' with the full model can inform our understanding of other 2L1S events with incomplete light-curve coverage.

Therefore, for the remainder of this section, we will exclude the MOA data on the peak.  These data will then be incorporated in Section \ref{3L1S}.

\subsection{Static Binary Lens Model}\label{sec:BL}

The 2L1S model requires three additional parameters $(s, q, \alpha)$, which are respectively the separation of the two lens bodies scaled to $\thetae$, the mass ratio between these bodies, and the angle of the source trajectory relative to the binary axis. For modeling, we use the advanced contour integration code \citep{Bozza2010,Bozza2018}, \texttt{VBBinaryLensing}\footnote{\url{http://www.fisica.unisa.it/GravitationAstrophysics/VBBinaryLensing.htm}}. We initially carry out a sparse grid searches for the parameters ($\log s, \log q, \alpha$). The grid consists of 21 values equally spaced between $-1.0 \leq \log s \leq 1.0$, 20 values equally spaced between $0^{\circ}\leq \alpha < 360^{\circ}$, and 61 values equally spaced between $-6.0 \leq \log q \leq 0.0$. For each set of ($\log s, \log q, \alpha$), we find the minimum $\chi^2$ by Markov chain Monte Carlo (MCMC) $\chi^2$ minimization using the \texttt{emcee} ensemble sampler \citep{emcee}, with fixed $\log q$, $\log s$ and free $t_0, u_0, \te, \rho, \alpha$. We identify one local minimum at $(\log s,\log q)\simeq (0.0, -5.1)$, similar to the case of \cite{OB190960}. We thus conduct a similar dense grid search as \cite{OB190960} that consists of 51 values equally spaced between $-0.02 \leq \log s \leq 0.03$ and 41 values equally spaced between $-6.0 \leq \log q \leq -4.0$. Often, such a grid search yields two local minima at $s > 1$ and $s < 1$ \citep[e.g.,][]{KB190842,OB190960}, which must then be individually further explored and compared. However, in the present case, there is only one local minimum at $s < 1$, while the $s > 1$ model is disfavored by $\Delta\chi^2 > 900$. We refine this minimum by allowing all parameters to vary. The parameters with their $68\%$ uncertainty range from the MCMC are shown in Table \ref{parm1}, and the fit and residuals are shown in Figure~\ref{lc1}. The very low mass ratio $q \sim 10^{-5}$ indicates that the companion is a very-low-mass planet.

\subsection{Microlens Parallax Model}\label{sec:parallax}

Even without detailed analysis, the results for the static model, that are listed in Table \ref{parm1}, imply a large (and so potentially measurable) microlens parallax \citep{Gould1992, Gould2000}, 
\begin{equation}\label{eqn:2}
    \bm{\pi}_{\rm E} \equiv \frac{\pi_{\rm rel}}{\thetae} \frac{\bm{\mu}_{\rm rel}}{\mu_{\rm rel}}.
\end{equation}
According to $\theta_*$ and the blended light in Section \ref{lens}, $\thetae = \theta_*/\rho \geq 1.68~{\rm mas}$ and the lens light $I_{\rm L} \geq 18.9$ at $3\sigma$ level. The two values correspond roughly to an $M \sim 0.5 M_{\odot}$ at a distance of 1.2 kpc. Thus, we can expect\footnote{One exception to this limit would be if the lens were a massive remnant.}
\begin{equation}\label{eqn:3}
    \pie = \frac{\thetae}{\kappa M_{\rm L}} \geq 0.41.
\end{equation}
Moreover, given the long Einstein timescale, $\te \sim 103$ days, the projected velocity on the observer plane, $\tilde v \equiv {\rm au}/(\pie \te) \la 42~{\rm km~s}^{-1}$ is close to the changes of Earth's velocity over the course of the event, taking account of which could impact other parameters as well.

Therefore, it is essential to include microlens-parallax effects in the fit. We fit the annual microlens-parallax effect by introducing two additional parameters $\pi_{\rm E,N}$ and $\pi_{\rm E,E}$, the North and East components of \bm{$\pie$} in equatorial coordinates \citep{Gouldpies2004}. We also fit $u_{0} > 0$ and $u_{0} < 0$ models to consider the ``ecliptic degeneracy'' \citep{Jiang2004, Poindexter2005}. Table \ref{parm1} shows the results of fitting the light curve with the microlens parallax effect. Because the annual parallax effect can be degenerate with the effects of lens orbital motion \citep{MB09387,OB09020}, we also introduce two linearized parameters ($ds/dt, d\alpha/dt$), the instantaneous changes in the separation and orientation of the two lens components defined at $t_0$. We restrict the MCMC trials to $\beta < 0.8$, where $\beta$ is the ratio of projected kinetic to potential energy \citep{OB050071D}
\begin{equation}\label{eq:orbit}
    \beta \equiv \left| \frac{{\rm KE}_{\perp}}{{\rm PE}_{\perp}} \right| = \frac{\kappa M_{\odot} {\rm yr}^2}{8\pi^2}\frac{\pie}{\thetae}\gamma^2\left(\frac{s}{\pie + \pis/\thetae}\right)^3; \qquad \bm{\gamma} \equiv \left(\frac{ds/dt}{s}, \frac{d\alpha}{dt}\right),
\end{equation}
where we adopt the source parallax $\pis = 0.128$ mas based on the mean distance to clump giant stars in this direction \citep{Nataf2013}. See Table \ref{parm1} for the results. We find that the addition of lens orbital motion effect provides improvements of $\Delta\chi^2 = 2.4~{\rm and}~5.4$ for the $u_0 > 0$ and $u_0 < 0$ solution, respectively, and $\bm{\pi}_{\rm E}$ is basically the same compared to the ``parallax'' model. 

Although the angular Einstein radius $\thetae$ estimated from the parallax modeling is smaller than the value from the static model, the resulting parallax for 2L1S is still strongly inconsistent with the constraint of the blended light at about $3\sigma$. We will further discuss the implication of the 2L1S results in Section \ref{2L1S_imp}.

\subsection{Binary-Source (1L2S) Model}\label{sec:BS}
In some cases, planetary (2L1S) light curves can be imitated by binary-source (1L2S) events \citep{Gaudi1998}.  We do not expect that this will be case for \event\ because the planetary anomaly is mainly characterized by sharp changes in slope, rather than a smooth short-lived bump. Nevertheless, as a matter of due diligence, we search for such models including both microlens-parallax and microlens-xarallap effects \citep{Griest1992,HanGould1997,Poindexter2005}. We find that while the introduction of a second source yields a huge improvement with respect to the 1L1S model with $\Delta\chi^2 = \chi^2({\rm 1L1S}) - \chi^2({\rm 1L2S}) > 10000$, the 1L2S model still does not compete with the 2L1S model with $\Delta\chi^2 = \chi^2({\rm 1L2S}) - \chi^2({\rm 2L1S}) > 400$.

%% file: 3L1S.tex
While the 2L1S models described in Section \ref{2L1S} fit the non-MOA-peak data very well, they completely fail to explain the features of the MOA data in the peak region.  Moreover, a 2L1S grid search that includes all the data fails to return any model that even approximately traces the data over the peak. See Figure \ref{lc1}. We therefore conduct a 3L1S grid search. Relative to static 2L1S models, 3L1S models have three additional parameters, $(s_3, q_3, \psi)$.  These are, respectively, the normalized separation of the third body from the primary, the mass ratio of the third body to the primary, and the angle of the second from the third body, as seen from the primary. Note that, to avoid confusion, we rename $(s, q)\rightarrow (s_2, q_2)$.

\subsection{3L1S Static Models}\label{3L1S_static}

We begin by conducting a grid search for static 3L1S solutions that is analogous to the one carried out previously for 2L1S solutions, but is substantially more computationally intensive. In a grid search (whether 2L1S or 3L1S), the lens geometry is held fixed at each grid point. However, for 2L1S, the geometry is specified by just two parameters, $(s_2, q_2)$, whereas for 3L1S, five geometric parameters are required, $(s_2, q_2, s_3, q_3, \psi)$. To reduce the grid of geometries from five to three dimensions, we consider a $(70 \times 70 \times 180)$ grid in $(s_3, q_3, \psi)$.  We hold $(s_2, q_2)$ fixed at the best-fit 2L1S model. We seed the remaining five parameters, $(t_0, u_0, \te, \rho, \alpha)$, at the best-fit 2L1S model, and we then allow these to vary. We initially consider only close models for the third body, i.e., $s_3 < 1$. For the grid-search phase, which relies on fixed geometries, we apply the map-making technique of \citet{OB04343} to evaluate the magnifications. This grid search yields only one local region of candidate solutions. 

We then seed an additional MCMC with the best grid point from this region and allow all 10 parameters to vary. Because the geometry now varies with each
step in the MCMC, we use the adaptive-image inverse-ray-shooting technique to evaluate the magnifications. The resulting parameters are shown in Table \ref{parm2}, and the model light curve is compared to the data in Figure \ref{lc1}. The corresponding caustic structure in the upper panel of Figure \ref{cau} shows that \event\ is a classic case of ``caustic factorization''. The caustic is nearly the superposition of two well-known caustic types: a large resonant caustic associated with the planet, and a smaller, nearly Chang-Refsdal \citep{CR}, caustic associated with the third body. As is often the case, the two caustic structures ``interact'' and become intertwined where they overlap. Because the caustic factors and $q_3 \ll 1$, it is straightforward to guess the alternate wide $(s_3 > 1)$ solution according to the prescription of \citet{Griest1998}: $s_3 \rightarrow s_3^{-1}$. We seed this guess into an MCMC to yield the alternate wide solution, whose parameters are given in Table \ref{parm2} and whose geometry is shown in the lower panel of Figure \ref{cau}. The most striking difference between the 3L1S close and wide solutions is that $s_{2,\rm close} = 0.99914 \pm 0.00011$, while $s_{2,\rm wide} = 0.96882 \pm 0.00018$, which appears to be a ``$100\sigma$'' difference. We address this issue in Appendix \S~\ref{x2}.

\subsection{3L1S Parallax-only Models}

As discussed in Section~\ref{sec:parallax}, the microlens-parallax parameters $\bm{\pi}_{\rm E}$ (Equation (\ref{eqn:2})) can be degenerate with the orbital-motion parameters $\bm{\gamma}$ (Equation~(\ref{eq:orbit})). Hence, both should be considered together. However, as in that section, we proceed step-by-step, in part due to the increasing computational load as more parameters are introduced, and therefore the importance of understanding which are really necessary.

When only $\bm{\pi}_{\rm E}$ is added to the 3L1S static model, there are 12 parameters. As was the case for 2L1S, adding parallax to the fit results in doubling the number of solutions, i.e., there is a $\pm u_0$ pair of solutions for each of the close and wide solutions found in Section \ref{3L1S_static}. Hence there are four solutions altogether. The resulting parameters and $\chi^2$ values are given in Table \ref{parm2}. It is found that including parallax significantly improves the fit by $\Delta\chi^2 > 130$. The wide solutions are disfavored by $\Delta\chi^2 > 17$. Between the two close solutions, the $u_0 > 0$ solution is significantly favored by $\Delta\chi^2 \sim 12$. We note that the magnitude of $\bm{\pi}_{\rm E}$ for 3L1S is substantially larger than the value estimated from the 2L1S modeling regardless of the lens-orbital effect. Figure \ref{cum} shows the cumulative distribution of $\Delta\chi^2 = \chi^2(\rm static) - \chi^2(\rm parallax)$ for the four solutions. Overall, $\Delta\chi^2$ grows steadily over time, giving credence to the parallax measurement. An important feature of this diagram is that the contribution to $\Delta\chi^2$ during the short time interval starting from one day before the peak and ending two days after the peak is about $40\%$ of the total $\Delta\chi^2$. By contrast, one normally expects the parallax signal to be dominated by the wings of the light curve. This time interval is essentially the duration of contact with the planetary caustic, in particular as the source ``rides the caustic'' for two days after the peak. This gives a plausible explanation for the sensitivity of $\bm{\pi}_{\rm E}$ to the near-peak region of the light curve. Hence, it is essential to include orbital motion. 


\subsection{3L1S Parallax and Planet Orbital Motion}

We now include both $\bm{\pi}_{\rm E}$ and $\bm{\gamma}$ for the planet in the 3L1S fit, for a total of 14 chain parameters. We show the parameters of this fit in the Table \ref{parm3}. It is found that including planet orbital motion significantly changes $\bm{\pi}_{\rm E}$ in both magnitude and direction, and $|\pie|$ of the $u_0 < 0$ solution is $\sim 1.8$ times greater than that of the $u_0 > 0$ solution. The close $u_0 < 0$ solution has the best fit to the observed data, while other solutions are only disfavored by $\Delta\chi^2 < 7$. Thus, we cannot exclude any solution from the light-curve analysis. The ratio $\beta$ of projected kinetic to potential energy is well measured, and all of the solutions have $\beta < 0.1$ at $3\sigma$. This relatively low value of the ratio suggests that the planet and the host may be aligned along the line of sight.

\subsection{Possible Orbital Motion of the Third Body}

The third body (with a brown-dwarf-like mass ratio $q_3 \sim 0.05$) must also undergo orbital motion. In the wide solutions, the period would be of order 100 years, implying that orbital motion of the third body would not affect the lensing light curve. For the close solutions, the period would be of order 50 days, and thus its orbital motion could affect the light curve. Nevertheless, we do not attempt to model orbital motion of the third body for several reasons. First, the duration of its pronounced perturbation ($\sim 0.3\,$days over the peak) is 10 times shorter than for the duration of the planetary signal, and its impact on the light curve is quadratic in the duration. Second, if there were clear prospects of a scientifically important result, such work would be warranted, but there are no such prospects (see Section \ref{future}). Finally, the results of 3L1S parallax + planet orbital motion required about two weeks of computations with 400 processors. The already prodigious use of computer time (which scales $\propto (n/2)!$ where $n$ is the number of chain parameters) would increase by a factor eight. We therefore decline to pursue this aspect of the problem.

%% file: lens.tex
Normally, if the angular Einstein radius $\thetae$ and the microlens parallax $\pie$ are well measured, one can simply determine the lens total mass $M_{\rm L}$ and the lens distance $D_{\rm L}$ by \citep{Gould1992, Gould2000}
\begin{equation}\label{eq:mass}
    M_{\rm L} = \frac{\thetae}{{\kappa}\pie};\qquad D_{\rm L} = \frac{\mathrm{AU}}{\pie\thetae + \pi_{\rm S}}.
\end{equation}
However, in the present case, the lens system is very close ($D_{\rm L} \sim 1$~kpc), and the large symmetric errors of $\pie$ can lead to an asymmetric distribution in inferred lens distance. Hence, we conduct a Bayesian analysis to estimate the lens physical parameters in Section \ref{Baye}. Before doing so, we estimate $\theta_*$ by a color-magnitude diagram (CMD) analysis \citep{Yoo2004} in Section \ref{CMD}, in order to estimate the angular Einstein radius by $\thetae = \theta_*/\rho$. We also study the blended light in Section \ref{blend} to obtain constraints on the lens light. Finally in Section \ref{future}, we illustrate how future high-resolution photometric and spectroscopic observations would clarify the nature of the system.  

\subsection{Color-Magnitude Diagram (CMD)}\label{CMD}

Figure \ref{cmd} shows the CMD of stars from the OGLE-III catalog \citep{OGLEIII} located within a square region with one side length of $240^{\prime\prime}$ centered at the location of \event, together with the source position (blue) and the centroid of the red giant clump (red). We measure the centroid of the red giant clump as $(V - I, I)_{\rm cl} = (2.04 \pm 0.01, 15.55 \pm 0.02)$. For the intrinsic centroid of the red giant clump, we adopt $(V - I, I)_{\rm cl,0} = (1.06, 14.35)$ \citep{Bensby2013,Nataf2013}. This implies that $A_I = 1.20$ and $E(V - I) = 0.98$ toward this direction. For the source color, which is independent of any model, we get $(V - I)_{\rm S} = 1.82 \pm 0.01$ by regression of MOA $V$ versus $R$ flux as the source magnification changes and a calibration to the OGLE-III scale using the field-star photometry from the same reductions. Because the source apparent brightness slightly depends on the model, for simplicity, we explicitly derive results for $I_{\rm S} = 19.12$ and then present a scaling relation for different source magnitudes. From this procedure, we obtain the intrinsic color and brightness of the source as $(V - I, I)_{\rm S,0} = (0.84 \pm 0.03, 17.92 \pm 0.03)$, suggesting that the source is a mid-G type dwarf \citep{Bessell1988}. Using the color/surface-brightness relation of \cite{Adams2018}, we obtain 
\begin{equation}
    \theta_* = 0.943 \pm 0.047~\mu {\rm as}.
\end{equation}
where the 5\% error is given by Table 3 of \cite{Adams2018}. Then, for any particular model with source magnitude $I_{\rm S}$, one can infer $\theta_* = 0.943 \times 10^{-0.2(I_{\rm S} - 19.12)}$.

\subsection{The Blended Light}\label{blend}

For \event, the baseline object was detected by the OGLE-III survey, $(V, I)_{\rm base} = (20.70 \pm 0.08, 18.46 \pm 0.03)$. We further consider the statistical errors due to the mottled background from unresolved stars \citep{MB03037}. We follow the approach of \cite{OB180532} using the \texttt{GalSim} package \citep{GalSim} with the readout noise of \cite{OGLEIV}. We find $\sigma_I = 0.08$ mag and $\sigma_V = 0.012$ mag, and thus the baseline object has $(V, I)_{\rm base} = (20.70 \pm 0.14, 18.46 \pm 0.09)$, yielding the blended light of $(V - I, I)_{\rm B} = (3.1^{+1.0}_{-0.6}, 19.32 \pm 0.20)$. This value is consistent with the lens properties that are predicted by the microlensing light-curve and CMD analyses. For example, using the median $\thetae$ and $\pie$ values of the close $u_0 < 0$ solutions, the host mass $M_1 = 0.25~M_{\odot}$ and it would have rough intrinsic brightness and color of $M_I \sim 9.8$, $(V - I)_0 \sim 3.0$. Assuming an extinction curve with a scale height of 120 pc, it would have $(A_{I}, E(V - I))_{\rm L} = (0.38, 0.31)$ at the lens distance $D_{\rm L} = 0.74$~kpc. These corresponds to $(V - I, I)_{\rm L} \sim (3.3, 19.5)$, which is shown as the cyan point (``naive lens'') in Figure \ref{cmd} and is quite consistent with the blend. 

We also check the astrometric alignment between the source and the baseline object from KMTA imaging. We find that the baseline object lies $(0.18^{\prime\prime}, 0.01^{\prime\prime})$ west and south of the source. Because the source position is derived from difference image analysis on highly magnified images, the uncertainty in the source position ($\lesssim 0.01^{\prime\prime}$) is negligible relative to the error in the baseline position. We estimate the error of baseline position by the fractional astrometric error being equal to the fractional photometric error \citep{KB190842}, $\sigma_{\rm ast} = 0.39\sigma_{I}{\rm FWHM}$ = $0.09^{\prime\prime}$. Hence, the baseline object is astrometrically consistent with the source (and thus lens) at $2\sigma$ level. Thus, it is plausible that most or all of the blended light is due to the lens. 

The alignment between the source and the baseline object can be immediately checked (i.e., 2021 bulge season) by the {\it Hubble Space Telescope (HST)} or by ground-based adaptive optics (AO) mounted on large ground-based telescopes (e.g., Keck, Subaru). Even if the alignment was demonstrated, the blended light could in principle come from a stellar companion to either the source or the lens. However, if the alignment is $\lesssim 50~{\rm mas}$, an additional stellar companion to the lens would have generated significant deviations on the peak, so the confirmation that the blend is well aligned to the lens could probably rule out the lens-companion scenario. Because the source and the blended light have significantly different colors, the possibility of the source companion can be checked by a measurement of the astrometric offset in different bands \citep{OB03235_AO}. It is a priori unlikely that the blended light is primarily due to ambient stars that are unassociated with the event because of the low surface density of stars relative to the 180 mas offset. If high-resolution imaging nevertheless showed that this were the case, it would imply that the parallax was even larger and the lens was less massive and closer than the best-estimated values \footnote{Or the host could conceivably be a white dwarf.}.


\subsection{Bayesian Analysis}\label{Baye}

In the Bayesian analysis, we choose the log-normal initial mass function of \cite{Charbrier2003} as the mass distribution of the lens. For the bulge and disk stellar number density, we choose the model used by \cite{Zhu2017spitzer} and \cite{MB11262}, respectively. For the dynamical distribution of the disk lens, we assume the disk lenses follow a rotation of $240~{\rm km~s}^{-1}$ \citep{Reid2014} with the velocity dispersion used by \cite{KB180748}. For the source and bulge lens dynamical distributions, we examine a \emph{Gaia} CMD \citep{Gaia2016AA,Gaia2018AA} using the stars within $5'$ and derive the proper motion of ``clump'' stars ($16.4 < G < 17, 1.9 < B_p - R_p < 2.35$). We obtain (in the heliocentric frame) 
\begin{equation}
\langle\bm{\mu}_{\rm bulge}(\ell, b)\rangle = (-6.19, -0.36) \pm (0.15, 0.13)~\text{mas yr}^{-1},
\end{equation}
\begin{equation}
\sigma(\bm{\mu}_{\rm bulge}) = (2.78, 2.39) \pm (0.11, 0.09) ~\text{mas yr}^{-1}.    
\end{equation} 

We create a sample of $5 \times 10^9$ simulated events from the Galactic model. We weight each simulated event, i, by
\begin{equation}
    \omega_{{\rm Gal},i} = \Gamma_{i} \mathcal{L}_{i}(\te) \mathcal{L}_{i}(\thetae) \mathcal{L}_{i}(\pi_{\rm E, N}, \pi_{\rm E, E}),
\end{equation}
where $\Gamma_{i}\varpropto\theta_{{\rm E},i}\times\mu_{{\rm rel},i}$ is the microlensing event rate, $\mathcal{L}_{i}(\te)$, $\mathcal{L}_{i}(\thetae)$ and $\mathcal{L}_{i}(\pi_{\rm E, N}, \pi_{\rm E, E})$ are the likelihoods of its inferred parameters given the distributions of these quantities. In addition, we adopt the $3\sigma$ upper limit of the blended light, $I_{\rm B, limit} = 18.9$, as the upper limit of the lens flux. We then adopt the mass-luminosity relation of \cite{OB171130},
\begin{equation}
    M_I = 4.4 - 8.5\log{(\frac{M_{\rm L}}{M_{\odot}})},
\end{equation}
where $M_I$ is the absolute magnitude in the $I$ band, and reject trial events for which the lens properties obey 
\begin{equation}
    M_I + 5 \log{\frac{D_{\rm L}}{10 {\rm pc}}} + A_{I, D_{\rm L}} < I_{\rm B,limit},
\end{equation}
where $A_{I, D_{\rm L}}$ is the extinction at $D_{\rm L}$, which is derived by an extinction curve with a scale height of 120 pc.

In Table \ref{phy}, we list the angular Einstein radius, $\thetae$, the microlens parallax, $\pie$, the relative proper motion, $\mu_{\rm rel}$, the estimated masses of the individual lens components, $M_1$, $M_2$ and $M_3$, the distance to the lens, $D_{\rm L}$, and the projected separations to the host, $a_{\perp,2}$ and $a_{\perp,3}$ for all of the four 3L1S solutions from the Bayesian analysis. For both solutions, it is found that the host is probably an $M_1 \sim 0.3M_{\odot}$ M-type dwarf, the second body is an Earth-mass terrestrial planet at a projected separation of $a_{\perp,2} \sim 1.5$~au, and the third body is an object at the planet/brown-dwarf boundary at a projected separation of $a_{\perp,3} \sim 0.15$~au for the close solutions and $a_{\perp,3} \sim 15$~au for the wide solutions. The lens system is located in the Galactic disk with a distance of $D_{\rm L} \sim 1.0$~kpc.

\subsection{Characterizing the Lens System by Future Observations}\label{future}

The highly degenerate 3L1S solutions and their very different $\bm{\pi}_{\rm E}$, with large uncertainties, show that the characteristics of the lens system cannot be completely determined by our work. The ambiguous elements mainly include two aspects: the mass and distance of the lens system and the projected separation of the third body. They can be clarified by future high-resolution photometric and spectroscopic observations.

For the lens mass and distance, they can be unambiguously determined by conducting high-resolution AO imaging observations when the source and lens are resolved. \cite{OB120950} resolved the source and lens of OGLE-2012-BLG-0950 using Keck AO and the $\it HST$ when they were separated by $\sim$ 34~mas, in a case for which the source and lens had approximately equal brightness. In the present case, the blended light and source also have approximately equal brightness, so the lens and source can be resolved in 2024 provided that the lens contributes a significant part of the blended light. The lens-source resolution can yield the measurement of the lens flux and the lens-source relative proper motion vector, $\bm{\mu_{\rm rel}}$ \citep[e.g.,][]{Alock2001,Kozlowski2007,Batista2015}. The lens flux would provide an independent mass-distance relationship, and $\bm{\mu_{\rm rel}}$ would further constrain $\thetae$ (Equation \ref{eqn:1}) and $\pie$ (Equation \ref{eqn:2}) by its magnitude and direction, respectively. Together, these can significantly reduce the uncertainties of the physical lens parameters.

For the orbit of the third body, future spectroscopy observations to measure the radial velocity (RV) can determine whether the third body lies inside or outside the planetary orbit. Following the procedure of \cite{OB180740}, we estimate that the close solution has a RV amplitude $v\sin(i)$ of order $1~{\rm km}~{\rm s}^{-1}$ with a period of order $0.15$ yr, and the wide solution has a $v\sin(i)$ of order $100~{\rm m}~{\rm s}^{-1}$ with a period of order $100$ yr. These RV amplitudes and differences are big enough to be distinguished by high-resolution spectrometers on VLT/Espresso or future 30m telescopes. Note that if the lens indeed has comparable brightness to the source, then these observations can be made immediately, i.e., long before the lens and source separate on the sky, because the lens and source RVs likely differ by tens or hundreds of km/s.

%% file: dis.tex
\subsection{Inconsistency of the 2L1S Model}\label{2L1S_imp}

In Section \ref{2L1S}, we adopted the perspective that there are no data over peak, and we, in particular, asked how the event would have been analyzed and reported in the absence of such data. Of course, one obvious difference is that the data fit well with the 2L1S model and so there would have been no report of a ``third body''. But here we see that there would also have been a report of a ``major puzzle'' about this parallax inconsistency. The derived parallax values in Table \ref{parm1} ($\pie = 0.229 \pm 0.042$ for the $u_0 > 0$ solution or $\pie = 0.173 \pm 0.040$ for the $u_0 < 0$ solution) are strongly inconsistent with Equation (\ref{eqn:3}), which was derived from a combination of the well-measured value of $\thetae$ and the photometric contraints on blended light. Most likely, this would have been ``explained'' as due to ``probable low-level systematics in the KMTA and/or MOA data''. Perhaps there would have been some additional commentary about the possibility that the host was a white dwarf, which would enable evasion of the photometric constraints. Then, it is very likely that the final reported parameters would have been given as the static solution, while the physical parameters would have been derived from a Bayesian analysis that simply ignored the parallax measurement. And finally, it would have been pointed out that the apparent discrepancies could ultimately be resolved by future high-resolution follow-up observations.

While the scenarios laid out in the previous paragraph are necessarily somewhat speculative, we believe that the great majority of people familiar with the microlensing-planet literature would broadly agree with this assessment of what would have happened. We rehearse them here in detail because all of the ``explanations'', ``caveats'', etc., in the previous paragraph are completely wrong.  In fact, once the peak data are added back into the light curve and a third body is included in the modeling, the supposed ``tension'' surrounding Equation (\ref{eqn:3}) entirely disappears. We note that the problem of the impact of undetected ``third bodies'' on microlensing solutions was analyzed by \citet{Zhu2014ApJb}. While that study focused on the impact on the measured parameters of the detected planet, the present work shows that higher-order parameters can be affected as well.

\subsection{Two Paths for Very Low Mass-Ratio Planets}

Over the past two years, the previously empty ``tip'' of the $(\log s,\log q)$ diagram has been gradually populated by new discoveries. While KMT-2019-BLG-0842Lb lies just below the ``pile up'' associated with the \cite{KB170165} ``break''  (see Figure \ref{sq}), three other recent discoveries (KMT-2018-BLG-0029Lb, OGLE-2019-BLG-0960Lb, KMT-2020-BLG-0414Lb), with $q \sim (6.0, 4.5, 3.6)~\qe$, all lie well below it. All four of these recent discoveries were detected via resonant caustics.  Moreover, for three of these (all except KMT-2018-BLG-0029Lb), the source trajectory angle with respect to the binary axis was oblique. For two of the four (OGLE-2019-BLG-0960Lb and KMT-2020-BLG-0414Lb), followup observations played a major or dominant role in the detection and characterization.

Due to inhomogeneous (and difficult to model) selection, it is probably not possible to draw precise statistical conclusions  from this particular sample.  However, the broad features listed above tend to support the idea that there is not a strong break below Neptune mass ratios, as had been previously conjectured.
Rather, the conditions and lens geometries required for such detections are rare, including resonant caustic geometries and dense coverage over/near the peak of high-magnification events, often aided by oblique trajectories. In two cases, the dense coverage was provided by KMTNet high $(\Gamma = 1\,{\rm hr}^{-1})$, or very high $(\Gamma = 4\,{\rm hr}^{-1})$ 3-continent monitoring, and in the other two cases by dense 3-continent follow-up observations.

These characteristics suggest two paths forward for probing the very low-$q$ cold-planet population. First, as discussed by \cite{OB190960}, it should be possible to construct a rigorous statistical sample by systematic investigation of all KMTNet high-magnification events, where the threshold for ``high-magnification'' would be set at some definite value, like $A_{\rm max} > 20$. Each event with pipeline parameters near or above this boundary would require tender-loving care (TLC) reductions to determine if it were truly in the sample, and, if so, it would be subjected to a dense grid search for planets. Both KMT-2018-BLG-0029Lb and KMT-2019-BLG-0842Lb
would definitely be recovered by such a search, and it is possible that OGLE-2019-BLG-0960Lb would be recovered as well. These reductions are costly in human effort, but if restricted to high-magnification events, the project would be feasible.

Second, KMT-2020-BLG-0414Lb shows that follow-up observations of high-magnification events in low-cadence fields can detect the most extreme systems. It is challenging, but not impossible, to derive rigorous statistical conclusions from such a sample. For example, \citet{mufun} showed that their sample of 13 very high-magnification $(A_{\rm max} > 200)$ events was statistically well-grounded because they showed that the $\mu$FUN follow-up campaign randomly sampled half of the underlying OGLE-III $A_{\rm max} > 200$ events during 2004-2008. This achievement required prodigious human effort because, in the majority of cases, it was
not possible to determine which events would be very high magnification based on (often sparse) OGLE-III data alone. Hence, it was necessary to ``patrol'' a much larger set of OGLE events with 1.3m SMARTS telescope observations to even determine which events should be densely monitored.  And even choosing the events to be patrolled required several hours of human effort per day. This level of effort, for 1--2 planets per year was the main reason for abandoning this approach once massive surveys were underway.

However, with the advent of KMTNet's 3-continent survey, together with (beginning in full in 2019) its real-time Alert-Finder system \citep{KMTAF}, it has become possible to identify rising high-magnification events with little or no ``patrolling''. The LCO network has eight 1m robotic telescopes at the same sites as KMTNet and a reaction time of about 15 minutes, which allows the rapid necessary coverage for high-magnification events. $\mu$FUN is mainly composed of amateur volunteers and has different sites\footnote{(see \url{http://www.astronomy.ohio-state.edu/~microfun/microfun.html})} from the current surveys, which can provide targeted coverage for high-magnification events when the survey sites have poor weather. Indeed, our 2020 follow-up program was meant to be a pilot for such a patrol-free (or patrol-lite) high-magnification ($A \gtrsim 20$) planet search. This original idea was undermined by Covid-19, which resulted in the closure of two out of three KMTNet observatories for the bulk of the 2020 season. Substantial patrolling was therefore required in this season. For KMT-2020-BLG-0414Lb, the beginning of follow-up observations at ``$A_{\rm now} > 10$'' was actually the result of such patrolling. Nevertheless, the detection of KMT-2020-BLG-0414Lb shows that this approach is viable.

\subsection{Multiple Low Mass-Ratio Companions}

Including \event, 15 triple-lens events have been published. Among these, six lens systems contain two planets orbiting a star and seven consist of a planet in a binary-star system (for details, see Table 1 of \citealt{KB190797}). OGLE-2016-BLG-0613L \citep{OB160613} and KMT-2020-BLG-0414L are composed of a star, a low-mass brown dwarf and a planet\footnote{For OGLE-2016-BLG-0613L, there are two degenerate solutions (a planet + binary stars) disfavored by $\Delta\chi^2 \geq 10$.}. Among the eight systems containing a host and two low mass-ratio ($q < 0.1$) companions, five were detected by a high-magnification event: OGLE-2006-BLG-109 ($u_0 = 0.0035$, \citealt{OB06109,OB06109_Dave}), OGLE-2012-BLG-0026 ($u_0 = 0.0088$, \citealt{OB120026,OB120026_AO,OB120026_Zhu}), OGLE-2018-BLG-0532 ($u_0 = 0.0079$, \citealt{OB180532}), KMT-2019-BLG-1953 ($u_0 = 0.0007$, \citealt{KB191953}) and \event\ ($u_0 = 0.0007$)\footnote{Here we adopt the mean $u_0$ of the underlying 1L1S event for degenerate solutions.}. High-magnification events provide an efficient channel for multiple low mass-ratio companions because the source trajectory goes very close to the host, where each planet/BD induces distortions in the magnification profile via their central or resonant caustics \citep{Gaudi1998_high}. 


There have been only two microlensing studies about the occurrence rate of systems with multiple low mass-ratio companions. Using the one two-planet system OGLE-2006-BLG-109L among the 13 $\mu$FUN $A > 200$ sample, \cite{mufun} found that the frequency of solar-like systems is 1/6 for all planetary systems. Using the two multiplanetary systems OGLE-2006-BLG-109L and OGLE-2014-BLG-1722L, and the detection efficiency of the six-year MOA survey combined with the four-year $\mu$FUN follow-up observations, \cite{OB141722} estimated that $6\% \pm 2\%$ of stars host two cold giant planets. The detection of KMT-2020-BLG-0414L suggests that the new follow-up program for high-magnification events can also give an estimate of the occurrence rate of systems with multiple low mass-ratio companions. 

In addition, although the new follow-up program aims to observe events located in KMTNet $\Gamma \leq 1\,{\rm hr}^{-1}$ fields, it is still important to follow up very high-magnification events located in KMTNet $\Gamma = 4\,{\rm hr}^{-1}$ fields. First, a simulation of \cite{Zhu2014ApJ} shows that even KMTNet $\Gamma = 6\,{\rm hr}^{-1}$ cadence is not intensive enough to capture the subtle anomalies for very high-magnification events. This has been demonstrated by the two less secure multiplanetary events OGLE-2018-BLG-0532 and KMT-2019-BLG-1953, for which a cadence of $\Gamma = 4\,{\rm hr}^{-1}$ was only barely adequate to detect the second planet. Second, the survey observations could suffer from nonlinearity or saturation on the very bright peak of very high-magnification events. For \event, its $I = 11.1$ peak is too bright for the current surveys with their normal exposure time (60s for KMTNet with three 1.6m telescopes, 100-120s for OGLE with a 1.3m telescope and 60s for MOA with a 1.8m telescope). The LCO network and most of the $\mu$FUN telescopes have smaller apertures and more flexible exposure time, so the new follow-up program can provide supplementary coverage for the very bright peak of very high-magnification events in KMTNet $\Gamma = 4\,{\rm hr}^{-1}$ fields. 

\acknowledgments
W.Z. thank Subo Dong for fruitful discussions. W.Z., S.M. and X.Z. acknowledge support by the National Science Foundation of China (Grant No. 11821303 and 11761131004). Work by C.H. was supported by the grants of National Research Foundation of Korea (2019R1A2C2085965 and 2020R1A4A2002885). Work by JCY was supported by JPL grant 1571564. This research has made use of the KMTNet system operated by the Korea Astronomy and Space Science Institute (KASI) and the data were obtained at three host sites of CTIO in Chile, SAAO in South Africa, and SSO in Australia. This research uses data obtained through the Telescope Access Program (TAP), which has been funded by the TAP member institutes. The MOA project is supported by JSPS KAKENHI Grant Number JSPS24253004, JSPS26247023, JSPS23340064, JSPS15H00781, JP16H06287, and JP17H02871. Wei Zhu was supported by the Natural Sciences and Engineering Research Council of Canada (NSERC) under the funding reference \#CITA 490888-16.

\appendix

\section{The difference in $s_2$}\label{x2}
For the 3L1S model, in fact, what the light curve is sensitive to is not directly $s_2$, i.e., the angular separation between $M_2$ and $M_1$, scaled to $\thetae$. Rather, it is directly sensitive to ``$x_2$'', which is the offset between $M_2$ and the center of mass of the body(ies) interior to $s_2$, scaled to $\theta_{\rm E, interior}$, i.e., the Einstein radius associated with the mass interior to $s_2$. Hence, we do not expect the $s_2$ from the various solutions (in particular, the wide and close topologies) to be equal, but rather their associated $x_2$.

For example, for the wide 3L1S static solution, $M_1$ is the only body interior to $s_2$. Hence, its position is also the center of mass. However, $\theta_{\rm E, interior} = \thetae/\sqrt{1+q_3}$. Hence, 
\begin{equation}
    x_{2, \rm wide} = \sqrt{1+q_3}s_{2, \rm wide} =  0.99493 \pm 0.00011.
\end{equation}
For the corresponding close solution, $\theta_{\rm E, interior} = \thetae$. However, the center of mass of the material interior to $s_2$ is offset from $M_1$ by $\Delta x = s_3 q_3/(1+q_3)$ and thus
\begin{equation}
    x_{2,\rm close} = \sqrt{s_{2}^{2} - 2 s_2 \Delta x \cos{\psi} + \Delta x^2} = 0.99522 \pm 0.00010. 
\end{equation}
That is, as expected, $x_{2,\rm close} = x_{2,\rm wide}$ to within $2.8 \times 10^{-4}$.

%% file: table.tex
\begin{table}[ht]
    \renewcommand\arraystretch{1.2}
    \begin{centering}
    \caption{Data used in the analysis with corresponding data reduction method and rescaling factors}
    \begin{tabular}{c c c c c c c c c}
    \hline
    \hline
    Collaboration & Site & Filter & Coverage (${\rm HJD}^{\prime}$) & $N_{\rm data}$ & Reduction Method & $(k, e_{\rm min})$ for 2L1S & $(k, e_{\rm min})$ for 3L1S \\
    \hline
    KMTNet & SSO & $I$ & 8900.3 -- 9139.0 & 139 & pySIS$^1$ & (1.23, 0.010) & (1.11, 0.020) \\
    KMTNet & SSO & $V$ & 8931.2 -- 9130.9 & 21  & pyDIA$^2$ &  &  \\
    MOA & & Red & 9024.2 -- 9145.9 & 240 & \cite{Bond2001} & (1.74, 0.006) & (1.35, 0.020) \\
    MOA   &  & $V$ & 9022.0 -- 9112.9 & 22 & \cite{Bond2001} &   &   \\
    LCO & SSO & $i$ & 9038.9 -- 9047.2 & 53 & ISIS$^2$ & (1.08, 0.002) & (0.35, 0.010) \\
    LCO & SAAO & $i$ & 9038.3 -- 9055.5 & 107 & ISIS & (1.31, 0.001) & (1.05, 0.005) \\
    LCO & McDonald & $i$ & 9040.7 -- 9046.7 & 62 & ISIS & (1.03, 0.003) & (1.10, 0.005) \\
     CFHT & & $i$ & 9041.9 -- 9051.9 & 23 & ISIS & (3.25, 0.000) & (1.10, 0.020) \\
    $\mu$FUN & OPD06 & $I$ & 9040.4 -- 9045.7 & 105 & \texttt{DoPHOT}$^3$ & (1.15, 0.002) & (1.33, 0.000) \\
    $\mu$FUN & OPD16 & $I$ & 9038.5 -- 9049.6 & 861 & \texttt{DoPHOT} &  & \\
    $\mu$FUN & PEST  & unfiltered & 9042.2 -- 9051.0 & 98 & \texttt{DoPHOT} & (0.82, 0.016) & (0.71, 0.020) \\
    $\mu$FUN & Possum & unfiltered & 9041.8 -- 9052.0 & 143 & \texttt{DoPHOT} & (0.81, 0.027) & (0.87. 0.020) \\
    \hline
    \hline
    \end{tabular}
    \end{centering}
    \tablecomments{${\rm HJD}^{\prime} = {\rm HJD} - 2450000$. MOA $V$-band data are only used to determine the source color. KMT SSO $V$-band data are not used due to poor seeing. OPD16 data are not used for the analysis due to bleeding from a bright star. For the MOA data, the 41 points on the peak ($9040.78 < {\rm HJD}^{\prime} < 9041.20$) are exlucded in the 2L1S analysis.\\
    $^1$ \cite{pysis} \\
    $^2$ MichaelDAlbrow/pyDIA: Initial Release on Github, doi:10.5281/zenodo.268049\\
    $^3$ \cite{Alard1998,Alard2000,CFHT} \\
    $^4$ \cite{dophot}}
    \label{data}
\end{table}

\begin{table}[htb]
    \renewcommand\arraystretch{1.05}
    \centering
    \caption{Parameters for 2L1S Model}
    \begin{tabular}{c|c|c c|c c}
    \hline
    \hline
    Parameters &  Static & \multicolumn{2}{c|}{Parallax} & \multicolumn{2}{c}{Parallax + Orbital Motion} \\
      &   &  $u_0 > 0$ & $u_0 < 0$ & $u_0 > 0$ & $u_0 < 0$ \\
    \hline
    $\chi^2/dof$  & $948.6/904$ & $902.8/902$ & $921.0/902$ & $900.4/900$ & $917.4/900$ \\
    \hline
    $t_{0}$ (${\rm HJD}^{\prime}$) & 9041.0590 & 9041.0596 & 9041.0561 & 9041.0594 & 9041.0562 \\
                                          & 0.0009 & 0.0010 & 0.0011 & 0.0011 & 0.0012 \\
    $u_{0} (10^{-3})$  & 0.125 & 0.153 & $-$0.156 & 0.160 & $-$0.170 \\
                       & 0.014 & 0.015 & 0.016 & 0.018 & 0.019 \\
    $\te$ (days) & 102.9 & 93.5 & 91.6 & 95.9 & 95.7 \\
                  & 6.3 & 4.9 & 4.9 & 6.2 & 6.1 \\
    $s$  & 0.99591 & 0.99542 & 0.99521 & 0.99504 & 0.99454 \\
         & 0.00025 & 0.00025 & 0.00028 & 0.00052 & 0.00058 \\
    $q$ ($10^{-5}$) & 0.69 & 0.95 & 1.01 & 1.07 & 1.24 \\
                    & 0.13 & 0.15 & 0.17 & 0.22 & 0.25 \\
    $\alpha$ (rad) & 0.0177 & 0.0199 & $-$0.0205 & 0.0214 & $-$0.0232  \\
                   & 0.0011 & 0.0011 & 0.0012 & 0.0022 & 0.0024 \\
    $\rho$ ($10^{-4}$) & 4.10 & 5.10 & 5.39 & 5.29 & 5.80 \\
                       & 0.53 & 0.56 & 0.62 & 0.63 & 0.66 \\
    $\pi_{\rm E, N}$ & ... & 0.035 & 0.080 & 0.035 & 0.076 \\
                     & ... & 0.016 & 0.022 & 0.016 & 0.022 \\
    $\pi_{\rm E, E}$ & ... & 0.226 & 0.153 & 0.225 & 0.147 \\
                     & ... & 0.041 & 0.038 & 0.041 & 0.038 \\
    $ds/dt ({\rm yr}^{-1})$ & ... & ... & ... & 0.196 & 0.345 \\
                            & ... & ... & ... & 0.251 & 0.279 \\
    $d\alpha/dt ({\rm yr}^{-1})$ & ... & .... & ... & 0.017 & $-$0.021 \\
                                 & ... & .... & ... & 0.059 & $-$0.061 \\
    $I_{\rm S}$ & 19.209 & 19.104 & 19.075 & 19.132 & 19.126 \\
                & 0.070 & 0.059 & 0.060 & 0.072 & 0.072 \\
    \hline
    \hline
    \end{tabular}
    \tablecomments{${\rm HJD}^{\prime} = {\rm HJD} - 2450000$. Uncertainties are given in the second line for each parameter. $t_0$ represents the time of closest approach of the source to the lens mass center. $u_0$ is the closest distance of the source to the lens mass center.}
    \label{parm1}
\end{table}

\begin{table*}[htb]
    \renewcommand\arraystretch{1.01}
    \centering
    \caption{Parameters for 3L1S Static and Parallax Model}
    \begin{tabular}{c|c c|c c c c}
    \hline
    \hline
     Parameters & \multicolumn{2}{c|}{Static} &  \multicolumn{4}{c}{Parallax} \\
    \hline
     & Close & Wide & Close $u_0 > 0$ & Close $u_0 < 0$ & Wide $u_0 > 0$ & Wide $u_0 < 0$ \\
    \hline
    $\chi^2/dof$  & $1093.8/942$ & $1094.7/942$ & $944.6/940$ & $956.2/940$ & $964.8/940$ & $962.1/940$   \\
    \hline
    $t_{0}$ (${\rm HJD}^{\prime}$) & 9041.0375 & 9041.0362 & 9041.0360 & 9041.0359  & 9041.0365 & 9041.0357  \\
                                                & 0.0009 & 0.0006 & 0.0012 & 0.0009 & 0.0010 & 0.0010  \\
    $u_{0} (10^{-3})$  & 0.719 & 0.719 & 0.688 & $-$0.684 & 0.747 & $-$0.706 \\
                       & 0.015 & 0.007 & 0.019 & 0.006 & 0.013 & 0.008 \\
    $\te$ (days) & 91.3 & 90.3 & 94.6 & 95.3 & 88.3 & 93.0  \\
                    & 1.1 & 1.1 & 2.1 & 0.7 & 1.2 & 0.9  \\
    $s_2$  & 0.99914 & 0.96882 & 0.99827 & 0.99832 & 0.96849 & 0.97315  \\
           & 0.00011 & 0.00018 & 0.00019 & 0.00014 & 0.00053 & 0.00023 \\
    $x_2$  & 0.99522 & 0.99494 & 0.99511 & 0.99521 & 0.99453 & 0.99495  \\
           & 0.00012 & 0.00012 & 0.00019 & 0.00011 & 0.00018 & 0.00014 \\
    $q_2 (10^{-5})$ & 1.20 & 1.26 & 1.13 & 1.12 & 1.41 & 1.23  \\
                    & 0.05 & 0.04 & 0.08 & 0.02 & 0.06 & 0.03  \\
    $\alpha$ (rad) & 0.0257 & 0.0218 & 0.0251 & $-$0.0253 & 0.0230 & $-$0.0224  \\
                   & 0.0005 & 0.0002 & 0.0006 & 0.0002 & 0.0004 & 0.0003 \\
    $s_3$  & 0.0955 & 10.4633 & 0.1096 & 0.1108 & 10.3204 & 9.6413 \\
           & 0.0001 & 0.0002 & 0.0034 & 0.0026 & 0.0457 & 0.0366  \\
    $q_3$  & 0.0584 & 0.0546 & 0.0404 & 0.0395 & 0.0545 & 0.0453 \\
           & 0.0011 & 0.0002 & 0.0029 & 0.0021 & 0.0011 & 0.0005 \\
    $\psi$ (rad) &  $-$0.729 & $-$0.746 & $-$0.730 & 0.733 & $-$0.730 & 0.741  \\
                              & 0.008 & 0.005 & 0.014 & 0.010 & 0.013 & 0.011 \\
    $\rho (10^{-4})$ & 5.60 & 5.86 & 5.78 & 5.72 & 6.69 & 6.10  \\
                     & 0.15 & 0.13 & 0.27 & 0.08 & 0.18 & 0.12  \\
    $\pi_{\rm E, N}$ & ... & ... & 0.057 & 0.321 & $-$0.011 & 0.343 \\
                     & ... & ... & 0.042 & 0.035 & 0.035 & 0.045 \\
    $\pi_{\rm E, E}$ & ... & ... & 0.467 & 0.196 & 0.366 & 0.222  \\
                     & ... & ... & 0.073 & 0.059 & 0.056 & 0.067 \\
    $I_{\rm S}$ & 19.082 & 19.040 & 19.123 & 19.125 & 19.010 & 19.073 \\
                 & 0.014 & 0.014 & 0.025 & 0.007 & 0.015 & 0.009 \\
    \hline
    \end{tabular}
    \tablecomments{${\rm HJD}^{\prime} = {\rm HJD} - 2450000$. Uncertainties are given in the second line for each parameter. $t_0$ represents the time of closest approach of the source to the lens mass center. $u_0$ is the closest distance of the source to the lens mass center. $x_2$ is a derived quantity and is not fitted independently. See Appendix \S~\ref{x2} for the definition of $x_2$.}
    \label{parm2}
\end{table*}

\begin{table*}[htb]
    \renewcommand\arraystretch{1.01}
    \centering
    \caption{Parameters for 3L1S Model with Parallax and Planet Orbital Motion}
    \begin{tabular}{c|c c c c}
    \hline
    \hline
     Parameters & Close $u_0 > 0$ & Close $u_0 < 0$ &  Wide $u_0 > 0$ & Wide $u_0 < 0$ \\
    \hline
    $\chi^2/dof$  & $939.7/938$ & $938.0/938$ & $944.8/938$ & $939.5/938$ \\
    \hline
    $t_{0}$ (${\rm HJD}^{\prime}$) & 9041.0369 & 9041.0362 & 9041.0357 & 9041.0355  \\
                                       & 0.0010 & 0.0012 & 0.0012 & 0.0011 \\
    $u_{0} (10^{-3})$  & 0.652 & $-$0.688 & 0.693 & $-$0.720 \\
                       & 0.020 & 0.011 & 0.017 & 0.012 \\
    $\te$ (days) & 99.4 & 94.4 & 93.8 & 90.4 \\
                    & 2.4 & 1.1 & 2.0 & 1.5 \\
    $s_2$  & 0.99925 & 0.99896 & 0.97202 & 0.96835 \\
           & 0.00026 & 0.00035 & 0.00173 & 0.00060 \\
    $x_2$  & 0.99537 & 0.99521 & 0.99503 & 0.99496 \\
           & 0.00023 & 0.00011 & 0.00026 & 0.00027 \\
    $q_2 (10^{-5})$ & 1.01 & 1.13 & 1.14 & 1.21 \\
                    & 0.11 & 0.06 & 0.07 & 0.06 \\
    $\alpha$ (rad) & 0.0255 & -0.0269 & 0.0222 & $-$0.0233 \\
                   & 0.0010 & 0.0005 & 0.0006 & 0.0005 \\
    $s_3$  & 0.0895 & 0.0940 & 9.9713 & 10.5031 \\
           & 0.0040 & 0.0051 & 0.4197 & 0.0607 \\
    $q_3$  & 0.0616 & 0.0578 & 0.0476 & 0.0557 \\
           & 0.0038 & 0.0071 & 0.0038 & 0.0013 \\
    $\psi$ (rad) &  $-$0.726 & 0.731 & $-$0.737 & 0.740 \\
                     & 0.013 & 0.013 & 0.015 & 0.013 \\
    $\rho (10^{-4})$ & 5.33 & 5.78 & 5.92 & 6.20 \\
                     & 0.30 & 0.16 & 0.23 & 0.17 \\
    $\pi_{\rm E, N}$ & 0.389 & $-$0.345 & 0.480 & $-$0.552 \\
                     & 0.154 & 0.120 & 0.152 & 0.203 \\
    $\pi_{\rm E, E}$ & 0.173 & 0.669 & 0.220 & 0.809 \\
                     & 0.135 & 0.105 & 0.121 & 0.140 \\
    $ds/dt ({\rm yr}^{-1})$ & 0.008 & $-$0.022 & -0.034 & $-$0.091 \\
                            & 0.054 & 0.055 & 0.062 & 0.054 \\
    $d\alpha/dt ({\rm yr}^{-1})$ & 0.466 & $-$0.848 & 0.496 & $-$1.054 \\
                                 & 0.169 & 0.137 & 0.181 & 0.212 \\
    $\beta$  & 0.042 & 0.058 & 0.035 & 0.055 \\
             & 0.018 & 0.006 & 0.014 & 0.005 \\
    $I_{\rm S}$ & 19.173 & 19.121 & 19.084 & 19.042 \\
                 & 0.026 & 0.013 & 0.021 & 0.017 \\
    \hline
    \end{tabular}
    \tablecomments{${\rm HJD}^{\prime} = {\rm HJD} - 2450000$. Uncertainties are given in the second line for each parameter. $t_0$ represents the time of closest approach of the source to the lens mass center. $u_0$ is the closest distance of the source to the lens mass center. $\beta$ and $x_2$ are derived quantities and are not fitted independently. See Equation (\ref{eq:orbit}) and Appendix \S~\ref{x2} for the definitions of $\beta$ and $x_2$, respectively.}
    \label{parm3}
\end{table*}

\begin{table}[ht]
    \renewcommand\arraystretch{1.20}
    \setlength{\tabcolsep}{3.5pt}
    \centering
    \caption{Physical Parameters}
    \begin{tabular}{c c c c c}
    \hline
    \hline
    Solutions & Close $u_0 > 0$ & Close $u_0 < 0$ & Wide $u_0 > 0$ & Wide $u_0 < 0$ \\
    \hline
    $\theta_{\rm E}$ [mas] & $1.475_{-0.091}^{+0.087}$ & $1.586_{-0.074}^{+0.074}$ & $1.489_{-0.083}^{+0.077}$ & $1.574_{-0.060}^{+0.057}$ \\
    $\pie$ & $0.470_{-0.096}^{+0.106}$ & $0.715_{-0.105}^{+0.096}$ & $0.513_{-0.097}^{+0.114}$ & $0.840_{-0.142}^{+0.158}$ \\
    $M_1$ [$M_{\odot}$] & $0.364_{-0.057}^{+0.072}$ & $0.259_{-0.034}^{+0.041}$ & $0.340_{-0.053}^{+0.063}$ & $0.217_{-0.037}^{+0.045}$ \\
    $M_2$ [$M_{\oplus}$] & $1.21_{-0.22}^{+0.28}$ & $0.96_{-0.13}^{+0.15}$ & $1.28_{-0.21}^{+0.25}$ & $0.86_{-0.13}^{+0.19}$ \\
    $M_3$ [$M_J$] & $23.3_{-3.9}^{+4.7}$ & $15.4_{-2.5}^{+3.3}$ & $16.9_{-3.0}^{+3.4}$ & $12.5_{-2.1}^{+2.6}$ \\
    $D_{\rm L}$ [kpc] & $1.22^{+0.32}_{-0.23}$ & $0.80^{+0.12}_{-0.10}$ & $1.12^{+0.27}_{-0.20}$ & $0.69^{+0.13}_{-0.11}$ \\
    $a_{\perp, 2}$ [au] & $1.79_{-0.28}^{+0.36}$ & $1.26^{+0.19}_{-0.14}$ & $1.62_{-0.26}^{+0.30}$ & $1.04^{+0.19}_{-0.16}$ \\
    $a_{\perp, 3}$ [au] & $0.16^{+0.03}_{-0.03}$ & $0.12^{+0.02}_{-0.02}$ & $16.6^{+3.2}_{-2.7}$ & $11.4^{+2.1}_{-1.8}$ \\
    $\mu_{\rm hel, N}$ [${\rm mas\,yr^{-1}}$] & $4.92_{-0.81}^{+0.51}$ & $-2.82_{-0.82}^{+0.94}$ & $5.22_{-0.76}^{+0.49}$ & $-3.58_{-0.94}^{+1.41}$  \\
    $\mu_{\rm hel, E}$ [${\rm mas\,yr^{-1}}$] & $6.12_{-1.92}^{+1.76}$ & $12.07_{-1.21}^{+1.38}$ & $6.91_{-1.72}^{+1.58}$ & $13.13_{-1.30}^{+1.67}$ \\
    \hline
    \end{tabular}\\
    \label{phy}
\end{table}

%% file: figure.tex
\begin{figure}[htb] 
    \centering
    \includegraphics[width=0.80\columnwidth]{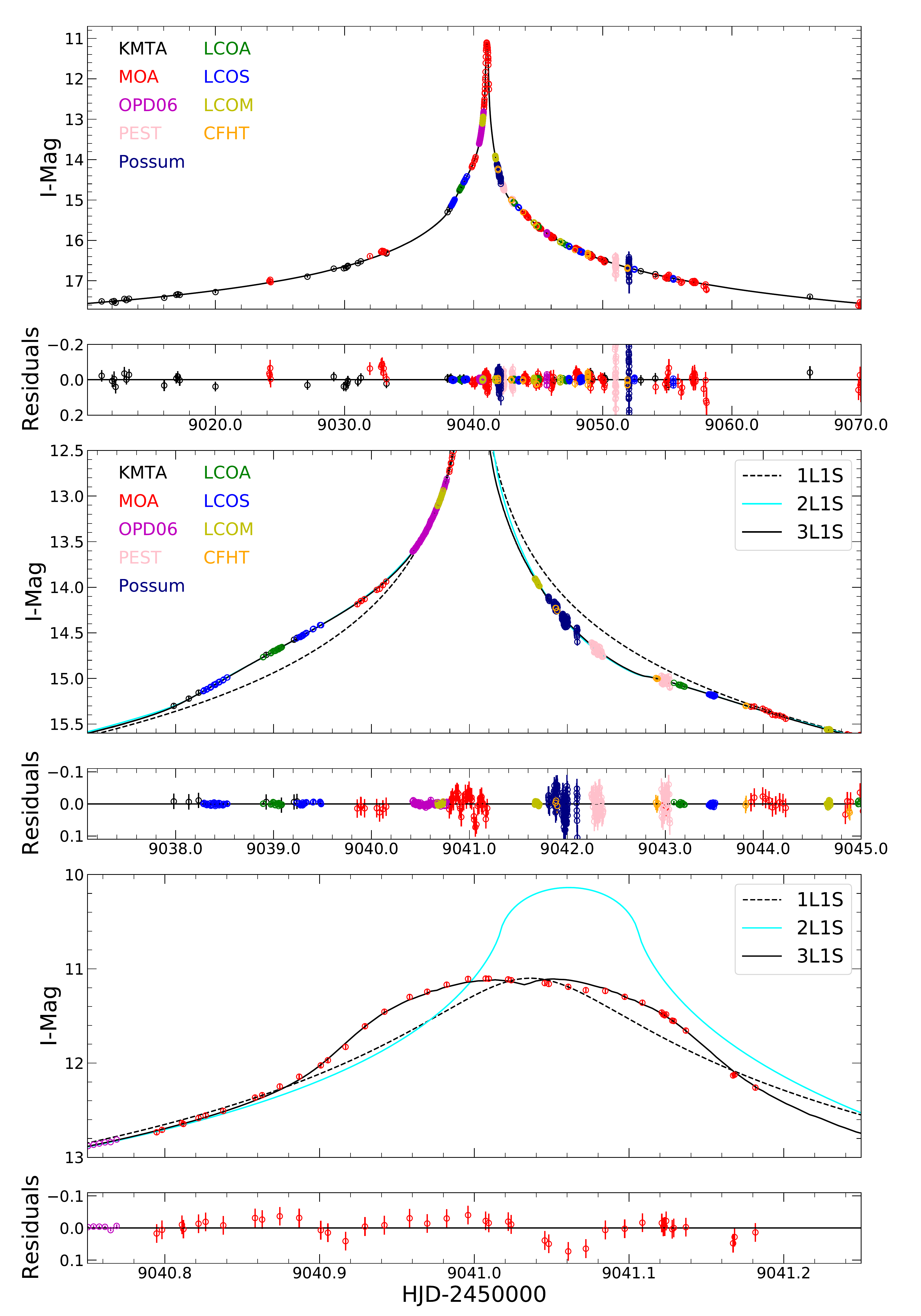}
    \caption{Light curve of \event\ with lensing models. The circles with different colors are the observed data points for different data sets. The black solid line is the best-fit 3L1S model using all the data, the cyan solid line is the best-fit 2L1S model excluding the MOA data on the peak, and the black dashed line is the 1L1S model derived using the same ($t_0, u_0, \te, \rho$) as the best-fit 3L1S model. The middle and bottom panels show a close-up of the main perturbations from the $q \sim 10^{-5}$ planet and the third body ($q \sim 0.05$), respectively.}
    \label{lc1}
\end{figure}



\begin{figure}[htb] 
    \centering
    \includegraphics[width=0.82\columnwidth]{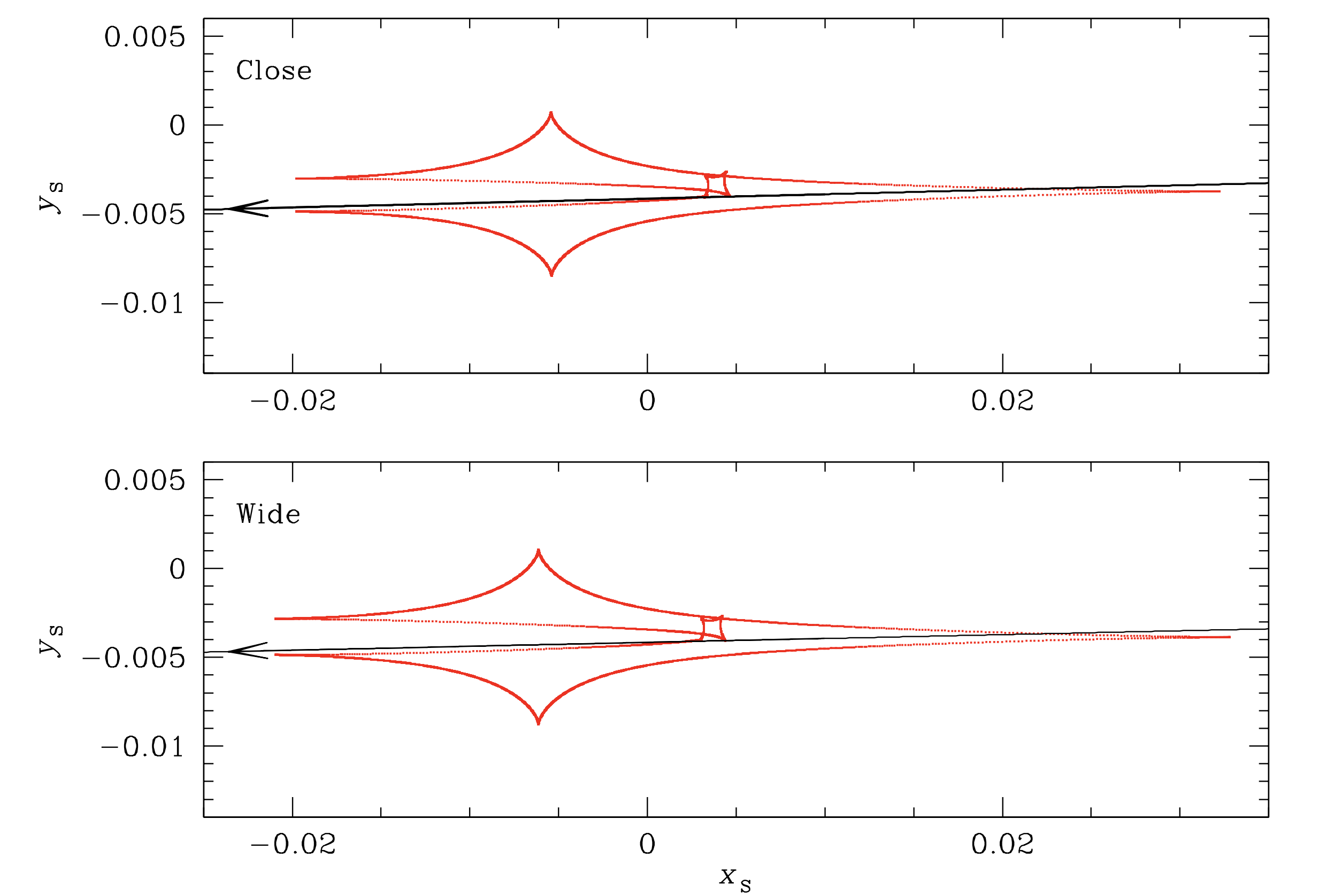}
    \caption{Geometries of the 3L1S Close (upper panel) and Wide (lower panel) models. In each panel, the red dashed line represents the caustic structure, the black solid line is the trajectory of the source, and the arrow indicates the direction of the source motion. The 3L1S caustic is nearly the superposition of a large 6-sided “resonant” caustic associated with the $q \sim 10^{-5}$ planet and a small quadrilateral caustic associated with the third body ($q \sim 0.05$).}
    \label{cau}
\end{figure}

\begin{figure}[htb] 
    \centering
    \includegraphics[width=0.65\columnwidth]{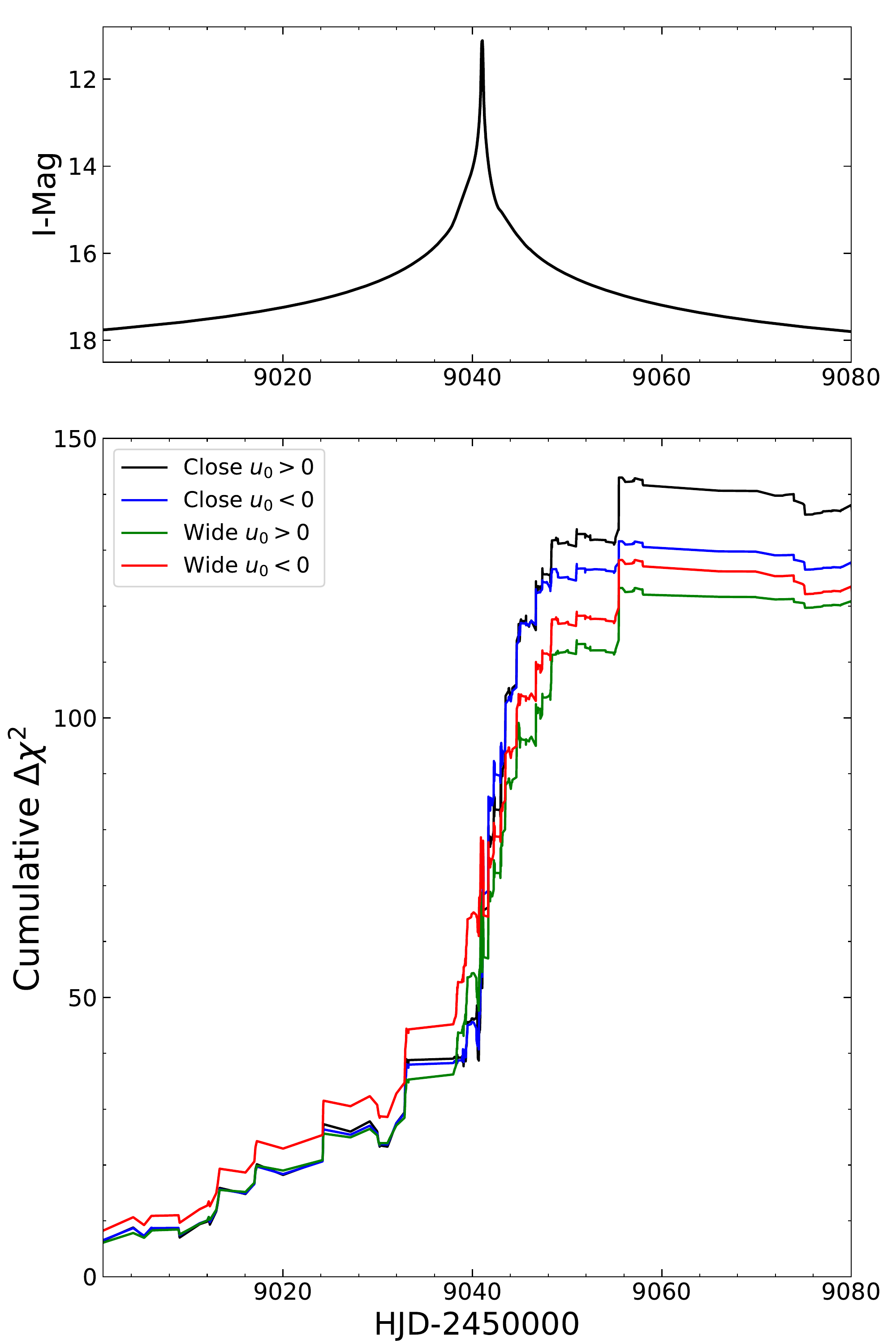}
    \caption{Cumulative distribution of $\Delta\chi^2 = \chi^2_{\rm static} - \chi^2_{\rm parallax}$ between the four 3L1S parallax solutions and the two 3L1S static solutions. Overall, $\Delta\chi^2$ grows steadily over time and does so steeply from one day before peak until two days after peak, when the source touches the planetary caustic. The upper panel shows the best-fit 3L1S parallax model.}
    \label{cum}
\end{figure}


\begin{figure}[htb] 
    \centering
    \includegraphics[width=0.85\columnwidth]{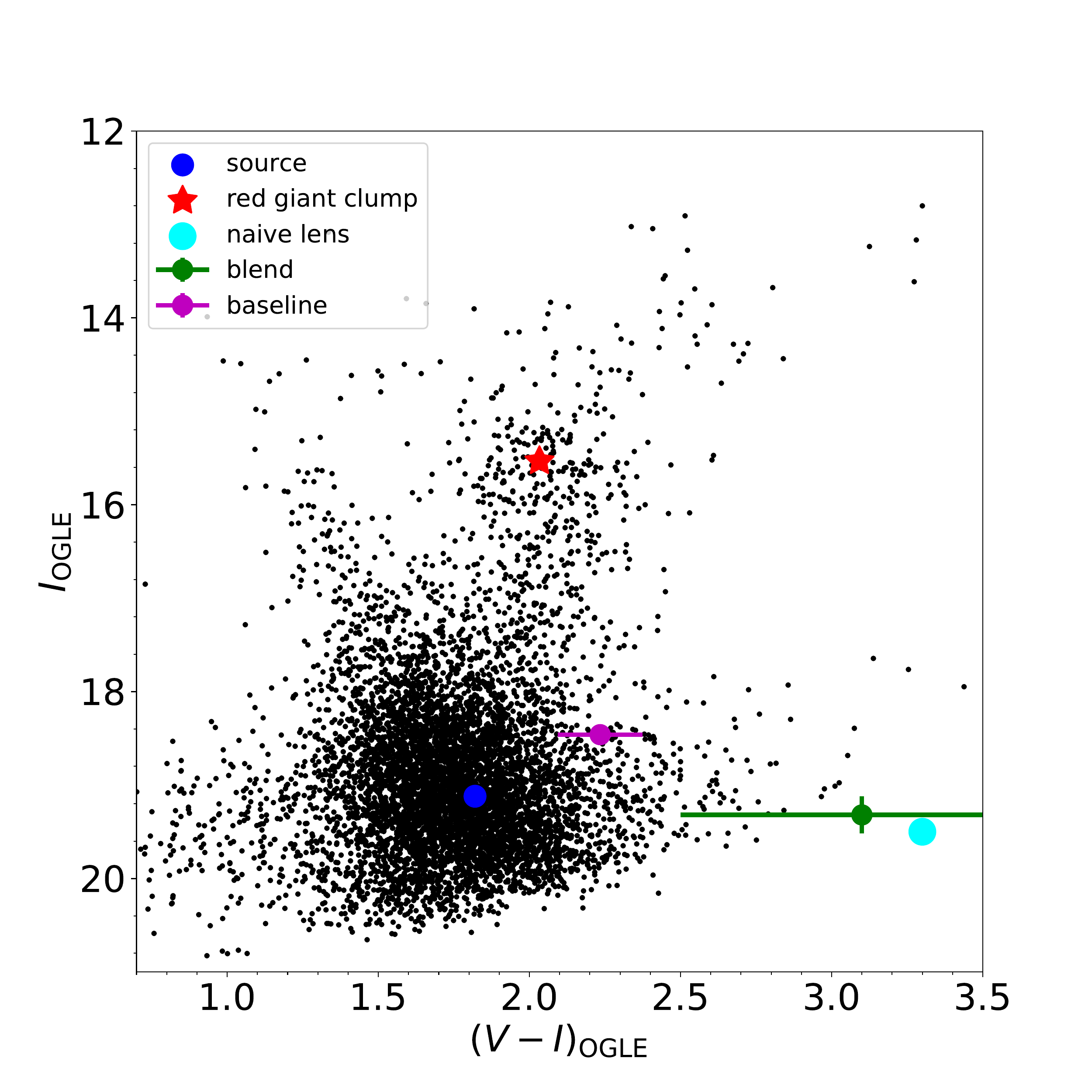}
    \caption{Color-magnitude diagram (CMD) for field stars in a $240^{\prime\prime}$ square centered on \event\ using the OGLE-III star catalog \citep{OGLEIII}. The red asterisk, blue dot, magenta dot and green dot represent the positions of the centroid of the red giant clump, the microlens source, the baseline object and the blended light, respectively. The cyan dot shows the position of a naive lens host with $M_{1} = 0.25M_{\odot}$ and $D_{\rm L} = 0.74$~kpc, estimated using the median $\thetae$ and $\pie$ values of the close $u_0 < 0$ solution.}
    \label{cmd}
\end{figure}


\begin{figure}
	\includegraphics[width=0.5\textwidth]{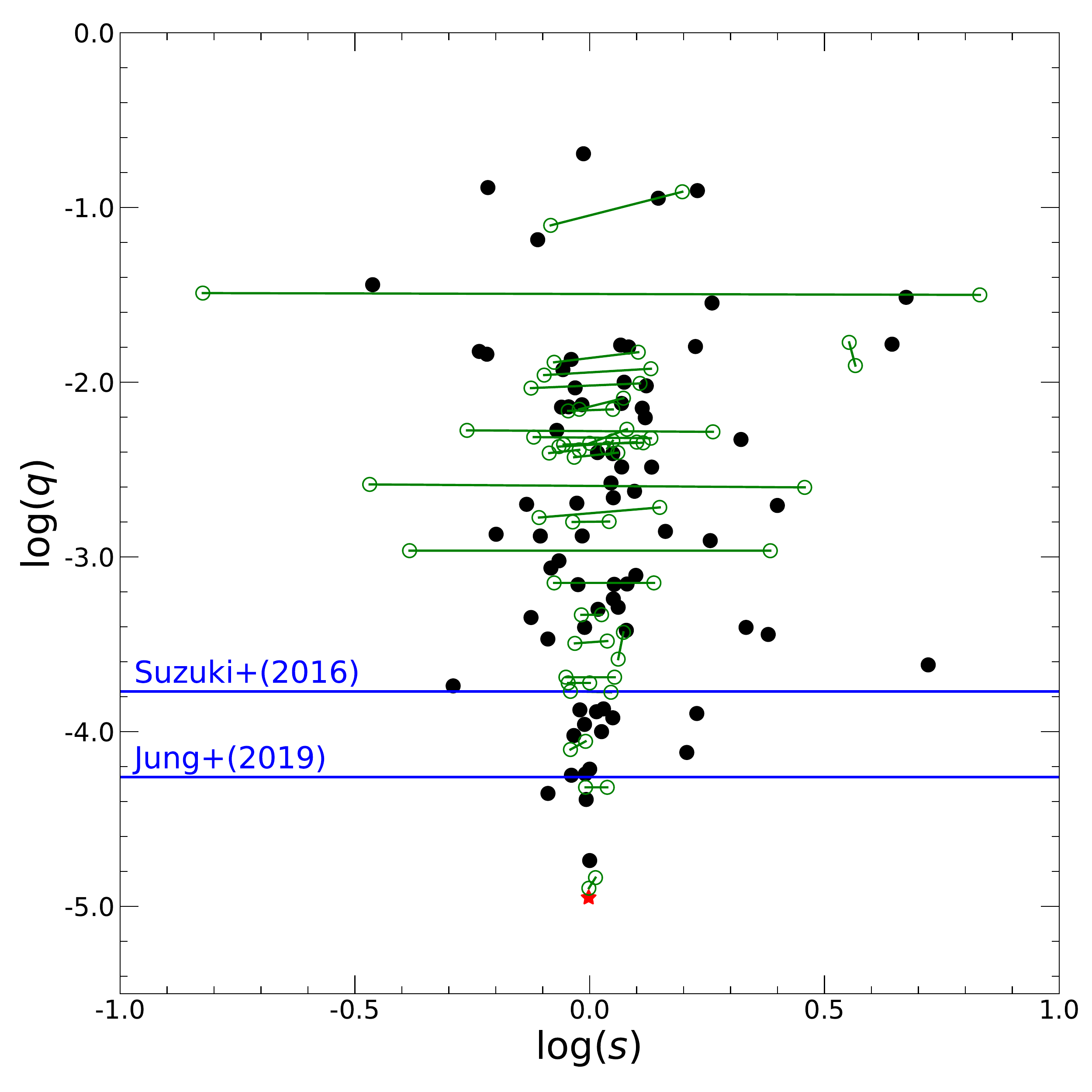}
	\includegraphics[width=0.5\textwidth]{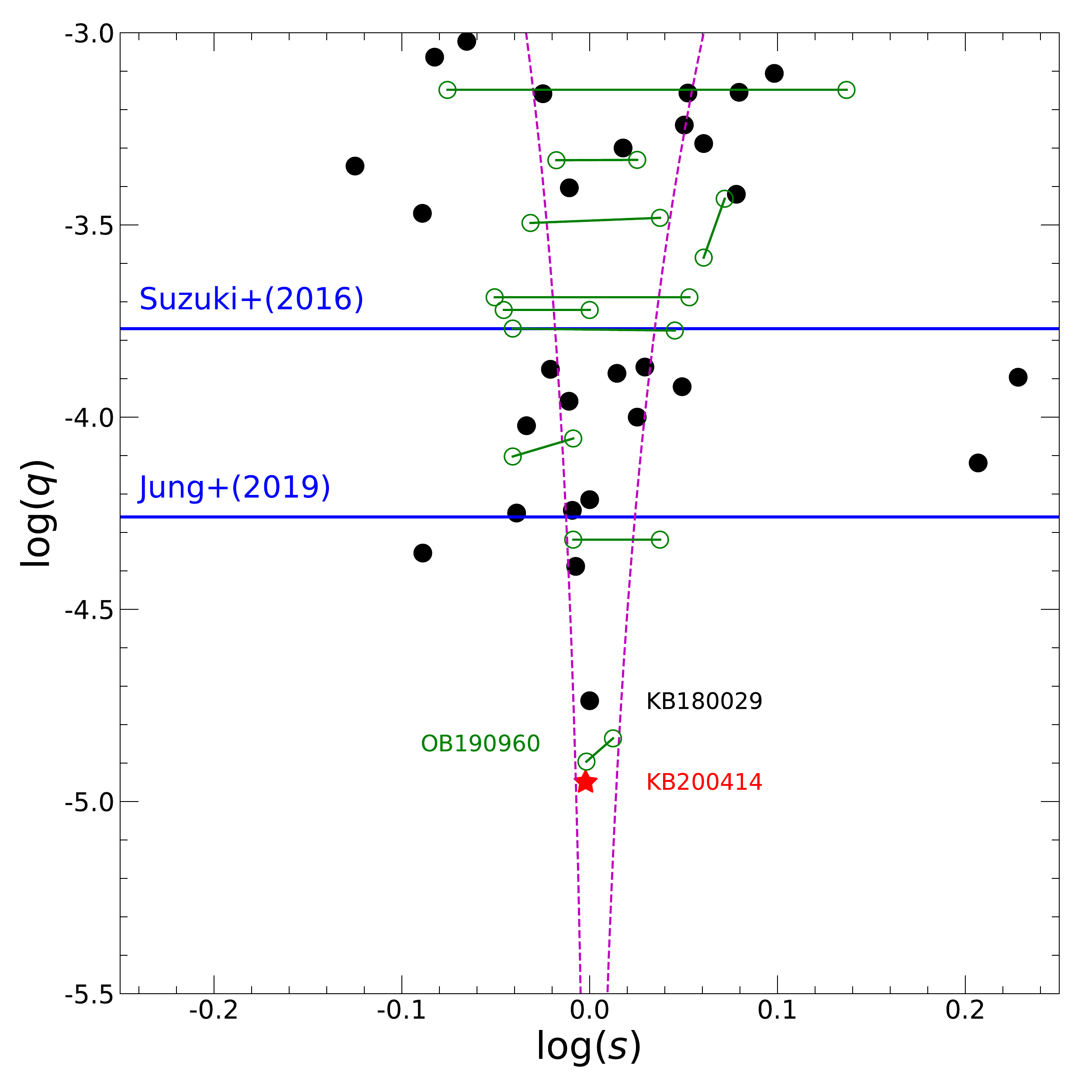}
	\caption{Microlensing parameters $(\log s,\log q)$ for planetary events ({\em Left:} all of the planets; {\em Right:} a close-up of $q < 10^{-3}$ planet), adapted from Figure 11 of \citet{OB190960}, but with the addition of KMT-2020-BLG-0414Lb (red asterisk, average of the median value of the four 3L1S solutions with parallax + planet orbital motion effects). The power-law ``breaks'' proposed by \cite{Suzuki2016} and \citet{KB170165} are indicated with the blue lines. Subsequently, there were three planetary events from 2018-2020 that lie well below the cluster of planets near the \citet{KB170165} ``break'' (KMT-2018-BLG-0029Lb, OGLE-2019-BLG-0960Lb, and KMT-2020-BLG-0414Lb), which are marked with text in the right panel. Together, these three recent discoveries ``fill out'' the previously empty region at the bottom of the triangular distribution. Solutions are considered to be ``unique'' (black points) if there are no competing solutions within $\Delta\chi^2<10$. Otherwise, they are shown by pairs of green open circles linked by a line segment. However, eight such pairs for which $q$ differs by more than a factor of two are excluded on the grounds that $q$ is not accurately measured. In the right panel, the two magenta dashed lines represent the boundaries between resonant and non-resonant caustics using the Equation (59) of \cite{Dominik1999}.}
	\label{sq}
\end{figure}

%% file: main.bbl
\begin{thebibliography}{}
\expandafter\ifx\csname natexlab\endcsname\relax\def\natexlab#1{#1}\fi

\bibitem[{{Abe} {et~al.}(2013){Abe}, {Airey}, {Barnard}, {Baudry}, {Botzler},
  {Douchin}, {Freeman}, {Larsen}, {Niemiec}, {Perrott}, {Philpott},
  {Rattenbury}, \& {Yock}}]{Abe2013}
{Abe}, F., {Airey}, C., {Barnard}, E., {et~al.} 2013, \mnras, 431, 2975

\bibitem[{{Adams} {et~al.}(2018){Adams}, {Boyajian}, \& {von
  Braun}}]{Adams2018}
{Adams}, A.~D., {Boyajian}, T.~S., \& {von Braun}, K. 2018, \mnras, 473, 3608

\bibitem[{{Alard}(2000)}]{Alard2000}
{Alard}, C. 2000, \aaps, 144, 363

\bibitem[{{Alard} \& {Lupton}(1998)}]{Alard1998}
{Alard}, C., \& {Lupton}, R.~H. 1998, \apj, 503, 325

\bibitem[{{Albrow} {et~al.}(2009){Albrow}, {Horne}, {Bramich}, {Fouqu{\'e}},
  {Miller}, {Beaulieu}, {Coutures}, {Menzies}, {Williams}, {Batista},
  {Bennett}, {Brillant}, {Cassan}, {Dieters}, {Dominis Prester}, {Donatowicz},
  {Greenhill}, {Kains}, {Kane}, {Kubas}, {Marquette}, {Pollard}, {Sahu},
  {Tsapras}, {Wambsganss}, \& {Zub}}]{pysis}
{Albrow}, M.~D., {Horne}, K., {Bramich}, D.~M., {et~al.} 2009, \mnras, 397,
  2099

\bibitem[{{Alcock} {et~al.}(2001){Alcock}, {Allsman}, {Alves}, {Axelrod},
  {Becker}, {Bennett}, {Cook}, {Drake}, {Freeman}, {Geha}, {Griest}, {Keller},
  {Lehner}, {Marshall}, {Minniti}, {Nelson}, {Peterson}, {Popowski}, {Pratt},
  {Quinn}, {Stubbs}, {Sutherland}, {Tomaney}, {Vandehei}, \&
  {Welch}}]{Alock2001}
{Alcock}, C., {Allsman}, R.~A., {Alves}, D.~R., {et~al.} 2001, \nat, 414, 617

\bibitem[{{Batista} {et~al.}(2015){Batista}, {Beaulieu}, {Bennett}, {Gould},
  {Marquette}, {Fukui}, \& {Bhattacharya}}]{Batista2015}
{Batista}, V., {Beaulieu}, J.-P., {Bennett}, D.~P., {et~al.} 2015, \apj, 808,
  170

\bibitem[{{Batista} {et~al.}(2011){Batista}, {Gould}, {Dieters}, {Dong},
  {Bond}, {Beaulieu}, {Maoz}, {Monard}, {Christie}, {McCormick}, {Albrow},
  {Horne}, {Tsapras}, {Burgdorf}, {Calchi Novati}, {Skottfelt}, {Caldwell},
  {Koz{\l}owski}, {Kubas}, {Gaudi}, {Han}, {Bennett}, {An}, {MOA
  Collaboration}, {Abe}, {Botzler}, {Douchin}, {Freeman}, {Fukui}, {Furusawa},
  {Hearnshaw}, {Hosaka}, {Itow}, {Kamiya}, {Kilmartin}, {Korpela}, {Lin},
  {Ling}, {Makita}, {Masuda}, {Matsubara}, {Miyake}, {Muraki}, {Nagaya},
  {Nishimoto}, {Ohnishi}, {Okumura}, {Perrott}, {Rattenbury}, {Saito},
  {Sullivan}, {Sumi}, {Sweatman}, {Tristram}, {von Seggern}, {Yock}, {PLANET
  Collaboration}, {Brillant}, {Calitz}, {Cassan}, {Cole}, {Cook}, {Coutures},
  {Dominis Prester}, {Donatowicz}, {Greenhill}, {Hoffman}, {Jablonski}, {Kane},
  {Kains}, {Marquette}, {Martin}, {Martioli}, {Meintjes}, {Menzies},
  {Pedretti}, {Pollard}, {Sahu}, {Vinter}, {Wambsganss}, {Watson}, {Williams},
  {Zub}, {FUN Collaboration}, {Allen}, {Bolt}, {Bos}, {DePoy}, {Drummond},
  {Eastman}, {Gal-Yam}, {Gorbikov}, {Higgins}, {Janczak}, {Kaspi}, {Lee},
  {Mallia}, {Maury}, {Monard}, {Moorhouse}, {Morgan}, {Natusch}, {Ofek},
  {Park}, {Pogge}, {Polishook}, {Santallo}, {Shporer}, {Spector}, {Thornley},
  {Yee}, {MiNDSTEp Consortium}, {Bozza}, {Browne}, {Dominik}, {Dreizler},
  {Finet}, {Glitrup}, {Grundahl}, {Harps{\o}e}, {Hessman}, {Hinse},
  {Hundertmark}, {J{\o}rgensen}, {Liebig}, {Maier}, {Mancini}, {Mathiasen},
  {Rahvar}, {Ricci}, {Scarpetta}, {Southworth}, {Surdej}, {Zimmer}, {RoboNet
  Collaboration}, {Allan}, {Bramich}, {Snodgrass}, {Steele}, \&
  {Street}}]{MB09387}
{Batista}, V., {Gould}, A., {Dieters}, S., {et~al.} 2011, \aap, 529, A102

\bibitem[{{Beaulieu} {et~al.}(2016){Beaulieu}, {Bennett}, {Batista}, {Fukui},
  {Marquette}, {Brillant}, {Cole}, {Rogers}, {Sumi}, {Abe}, {Bhattacharya},
  {Koshimoto}, {Suzuki}, {Tristram}, {Han}, {Gould}, {Pogge}, \&
  {Yee}}]{OB120026_AO}
{Beaulieu}, J.~P., {Bennett}, D.~P., {Batista}, V., {et~al.} 2016, \apj, 824,
  83

\bibitem[{{Bennett} {et~al.}(2006){Bennett}, {Anderson}, {Bond}, {Udalski}, \&
  {Gould}}]{OB03235_AO}
{Bennett}, D.~P., {Anderson}, J., {Bond}, I.~A., {Udalski}, A., \& {Gould}, A.
  2006, \apjl, 647, L171

\bibitem[{{Bennett} {et~al.}(2010){Bennett}, {Rhie}, {Nikolaev}, {Gaudi},
  {Udalski}, {Gould}, {Christie}, {Maoz}, {Dong}, {McCormick}, {Szyma{\'n}ski},
  {Tristram}, {Macintosh}, {Cook}, {Kubiak}, {Pietrzy{\'n}ski},
  {Soszy{\'n}ski}, {Szewczyk}, {Ulaczyk}, {Wyrzykowski}, {OGLE Collaboration},
  {DePoy}, {Han}, {Kaspi}, {Lee}, {Mallia}, {Natusch}, {Park}, {Pogge},
  {Polishook}, {{\ensuremath{\mu}}FUN Collaboration}, {Abe}, {Bond}, {Botzler},
  {Fukui}, {Hearnshaw}, {Itow}, {Kamiya}, {Korpela}, {Kilmartin}, {Lin},
  {Ling}, {Masuda}, {Matsubara}, {Motomura}, {Muraki}, {Nakamura}, {Okumura},
  {Ohnishi}, {Perrott}, {Rattenbury}, {Sako}, {Saito}, {Sato}, {Skuljan},
  {Sullivan}, {Sumi}, {Sweatman}, {Yock}, {MOA Collaboration}, {Albrow},
  {Allan}, {Beaulieu}, {Bramich}, {Burgdorf}, {Coutures}, {Dominik}, {Dieters},
  {Fouqu{\'e}}, {Greenhill}, {Horne}, {Snodgrass}, {Steele}, {Tsapras},
  {PLANET}, {RoboNet Collaborations}, {Chaboyer}, {Crocker}, \&
  {Frank}}]{OB06109_Dave}
{Bennett}, D.~P., {Rhie}, S.~H., {Nikolaev}, S., {et~al.} 2010, \apj, 713, 837

\bibitem[{{Bennett} {et~al.}(2014){Bennett}, {Batista}, {Bond}, {Bennett},
  {Suzuki}, {Beaulieu}, {Udalski}, {Donatowicz}, {Bozza}, {Abe}, {Botzler},
  {Freeman}, {Fukunaga}, {Fukui}, {Itow}, {Koshimoto}, {Ling}, {Masuda},
  {Matsubara}, {Muraki}, {Namba}, {Ohnishi}, {Rattenbury}, {Saito}, {Sullivan},
  {Sumi}, {Sweatman}, {Tristram}, {Tsurumi}, {Wada}, {Yock}, {MOA
  Collaboration}, {Albrow}, {Bachelet}, {Brillant}, {Caldwell}, {Cassan},
  {Cole}, {Corrales}, {Coutures}, {Dieters}, {Dominis Prester}, {Fouqu{\'e}},
  {Greenhill}, {Horne}, {Koo}, {Kubas}, {Marquette}, {Martin}, {Menzies},
  {Sahu}, {Wambsganss}, {Williams}, {Zub}, {PLANET Collaboration}, {Choi},
  {DePoy}, {Dong}, {Gaudi}, {Gould}, {Han}, {Henderson}, {McGregor}, {Lee},
  {Pogge}, {Shin}, {Yee}, {{\ensuremath{\mu}}FUN Collaboration},
  {Szyma{\'n}ski}, {Skowron}, {Poleski}, {Koz{\l}owski}, {Wyrzykowski},
  {Kubiak}, {Pietrukowicz}, {Pietrzy{\'n}ski}, {Soszy{\'n}ski}, {Ulaczyk},
  {OGLE Collaboration}, {Tsapras}, {Street}, {Dominik}, {Bramich}, {Browne},
  {Hundertmark}, {Kains}, {Snodgrass}, {Steele}, {RoboNet Collaboration},
  {Dekany}, {Gonzalez}, {Heyrovsk{\'y}}, {Kandori}, {Kerins}, {Lucas},
  {Minniti}, {Nagayama}, {Rejkuba}, {Robin}, \& {Saito}}]{MB11262}
{Bennett}, D.~P., {Batista}, V., {Bond}, I.~A., {et~al.} 2014, \apj, 785, 155

\bibitem[{{Bensby} {et~al.}(2013){Bensby}, {Yee}, {Feltzing}, {Johnson},
  {Gould}, {Cohen}, {Asplund}, {Mel{\'e}ndez}, {Lucatello}, {Han}, {Thompson},
  {Gal-Yam}, {Udalski}, {Bennett}, {Bond}, {Kohei}, {Sumi}, {Suzuki}, {Suzuki},
  {Takino}, {Tristram}, {Yamai}, \& {Yonehara}}]{Bensby2013}
{Bensby}, T., {Yee}, J.~C., {Feltzing}, S., {et~al.} 2013, \aap, 549, A147

\bibitem[{{Bessell} \& {Brett}(1988)}]{Bessell1988}
{Bessell}, M.~S., \& {Brett}, J.~M. 1988, \pasp, 100, 1134

\bibitem[{{Bhattacharya} {et~al.}(2018){Bhattacharya}, {Beaulieu}, {Bennett},
  {Anderson}, {Koshimoto}, {Lu}, {Batista}, {Blackman}, {Bond}, {Fukui},
  {Henderson}, {Hirao}, {Marquette}, {Mroz}, {Ranc}, \& {Udalski}}]{OB120950}
{Bhattacharya}, A., {Beaulieu}, J.~P., {Bennett}, D.~P., {et~al.} 2018, \aj,
  156, 289

\bibitem[{{Bond} {et~al.}(2001){Bond}, {Abe}, {Dodd}, {Hearnshaw}, {Honda},
  {Jugaku}, {Kilmartin}, {Marles}, {Masuda}, {Matsubara}, {Muraki}, {Nakamura},
  {Nankivell}, {Noda}, {Noguchi}, {Ohnishi}, {Rattenbury}, {Reid}, {Saito},
  {Sato}, {Sekiguchi}, {Skuljan}, {Sullivan}, {Sumi}, {Takeuti}, {Watase},
  {Wilkinson}, {Yamada}, {Yanagisawa}, \& {Yock}}]{Bond2001}
{Bond}, I.~A., {Abe}, F., {Dodd}, R.~J., {et~al.} 2001, \mnras, 327, 868

\bibitem[{{Bozza}(2010)}]{Bozza2010}
{Bozza}, V. 2010, \mnras, 408, 2188

\bibitem[{{Bozza} {et~al.}(2018){Bozza}, {Bachelet}, {Bartoli{\'c}}, {Heintz},
  {Hoag}, \& {Hundertmark}}]{Bozza2018}
{Bozza}, V., {Bachelet}, E., {Bartoli{\'c}}, F., {et~al.} 2018, \mnras, 479,
  5157

\bibitem[{{Chabrier}(2003)}]{Charbrier2003}
{Chabrier}, G. 2003, \pasp, 115, 763

\bibitem[{{Chang} \& {Refsdal}(1979)}]{CR}
{Chang}, K., \& {Refsdal}, S. 1979, \nat, 282, 561

\bibitem[{{Dominik}(1999)}]{Dominik1999}
{Dominik}, M. 1999, \aap, 349, 108

\bibitem[{{Dong} {et~al.}(2006){Dong}, {DePoy}, {Gaudi}, {Gould}, {Han},
  {Park}, {Pogge}, {MuFun Collaboration}, {Udalski}, {Szewczyk}, {Kubiak},
  {Szyma{\'n}ski}, {Pietrzy{\'n}ski}, {Soszy{\'n}ski}, {Wyrzykowski},
  {{\.Z}ebru{\'n}}, \& {OGLE Collaboration}}]{OB04343}
{Dong}, S., {DePoy}, D.~L., {Gaudi}, B.~S., {et~al.} 2006, \apj, 642, 842

\bibitem[{{Dong} {et~al.}(2009){Dong}, {Gould}, {Udalski}, {Anderson},
  {Christie}, {Gaudi}, {OGLE Collaboration}, {Jaroszy{\'n}ski}, {Kubiak},
  {Szyma{\'n}ski}, {Pietrzy{\'n}ski}, {Soszy{\'n}ski}, {Szewczyk}, {Ulaczyk},
  {Wyrzykowski}, {{$\mu$}FUN Collaboration}, {DePoy}, {Fox}, {Gal-Yam}, {Han},
  {L{\'e}pine}, {McCormick}, {Ofek}, {Park}, {Pogge}, {MOA Collaboration},
  {Abe}, {Bennett}, {Bond}, {Britton}, {Gilmore}, {Hearnshaw}, {Itow},
  {Kamiya}, {Kilmartin}, {Korpela}, {Masuda}, {Matsubara}, {Motomura},
  {Muraki}, {Nakamura}, {Ohnishi}, {Okada}, {Rattenbury}, {Saito}, {Sako},
  {Sasaki}, {Sullivan}, {Sumi}, {Tristram}, {Yanagisawa}, {Yock}, {Yoshoika},
  {PLANET/RoboNet Collaborations}, {Albrow}, {Beaulieu}, {Brillant}, {Calitz},
  {Cassan}, {Cook}, {Coutures}, {Dieters}, {Dominis Prester}, {Donatowicz},
  {Fouqu{\'e}}, {Greenhill}, {Hill}, {Hoffman}, {Horne}, {J{\o}rgensen},
  {Kane}, {Kubas}, {Marquette}, {Martin}, {Meintjes}, {Menzies}, {Pollard},
  {Sahu}, {Vinter}, {Wambsganss}, {Williams}, {Bode}, {Bramich}, {Burgdorf},
  {Snodgrass}, {Steele}, {Doublier}, \& {Foellmi}}]{OB050071D}
{Dong}, S., {Gould}, A., {Udalski}, A., {et~al.} 2009, \apj, 695, 970

\bibitem[{{Foreman-Mackey} {et~al.}(2013){Foreman-Mackey}, {Hogg}, {Lang}, \&
  {Goodman}}]{emcee}
{Foreman-Mackey}, D., {Hogg}, D.~W., {Lang}, D., \& {Goodman}, J. 2013, \pasp,
  125, 306

\bibitem[{{Gaia Collaboration} {et~al.}(2016){Gaia Collaboration}, {Prusti},
  {de Bruijne}, {Brown}, {Vallenari}, {Babusiaux}, {Bailer-Jones}, {Bastian},
  {Biermann}, {Evans}, \& et~al.}]{Gaia2016AA}
{Gaia Collaboration}, {Prusti}, T., {de Bruijne}, J.~H.~J., {et~al.} 2016,
  \aap, 595, A1

\bibitem[{{Gaia Collaboration} {et~al.}(2018){Gaia Collaboration}, {Brown},
  {Vallenari}, {Prusti}, {de Bruijne}, {Babusiaux}, {Bailer-Jones}, {Biermann},
  {Evans}, {Eyer}, \& et~al.}]{Gaia2018AA}
{Gaia Collaboration}, {Brown}, A.~G.~A., {Vallenari}, A., {et~al.} 2018, \aap,
  616, A1

\bibitem[{{Gaudi}(1998)}]{Gaudi1998}
{Gaudi}, B.~S. 1998, \apj, 506, 533

\bibitem[{{Gaudi} {et~al.}(1998){Gaudi}, {Naber}, \&
  {Sackett}}]{Gaudi1998_high}
{Gaudi}, B.~S., {Naber}, R.~M., \& {Sackett}, P.~D. 1998, \apjl, 502, L33

\bibitem[{{Gaudi} {et~al.}(2002){Gaudi}, {Albrow}, {An}, {Beaulieu},
  {Caldwell}, {DePoy}, {Dominik}, {Gould}, {Greenhill}, {Hill}, {Kane},
  {Martin}, {Menzies}, {Naber}, {Pel}, {Pogge}, {Pollard}, {Sackett}, {Sahu},
  {Vermaak}, {Vreeswijk}, {Watson}, \& {Williams}}]{Gaudi2002}
{Gaudi}, B.~S., {Albrow}, M.~D., {An}, J., {et~al.} 2002, \apj, 566, 463

\bibitem[{{Gaudi} {et~al.}(2008){Gaudi}, {Bennett}, {Udalski}, {Gould},
  {Christie}, {Maoz}, {Dong}, {McCormick}, {Szyma{\'n}ski}, {Tristram},
  {Nikolaev}, {Paczy{\'n}ski}, {Kubiak}, {Pietrzy{\'n}ski}, {Soszy{\'n}ski},
  {Szewczyk}, {Ulaczyk}, {Wyrzykowski}, {OGLE Collaboration}, {DePoy}, {Han},
  {Kaspi}, {Lee}, {Mallia}, {Natusch}, {Pogge}, {Park}, {{\ensuremath{\mu}}-Fun
  Collabortion}, {Abe}, {Bond}, {Botzler}, {Fukui}, {Hearnshaw}, {Itow},
  {Kamiya}, {Korpela}, {Kilmartin}, {Lin}, {Masuda}, {Matsubara}, {Motomura},
  {Muraki}, {Nakamura}, {Okumura}, {Ohnishi}, {Rattenbury}, {Sako}, {Saito},
  {Sato}, {Skuljan}, {Sullivan}, {Sumi}, {Sweatman}, {Yock}, {MOA
  Collaboration}, {Albrow}, {Allan}, {Beaulieu}, {Burgdorf}, {Cook},
  {Coutures}, {Dominik}, {Dieters}, {Fouqu{\'e}}, {Greenhill}, {Horne},
  {Steele}, {Tsapras}, {Planet Collaboration}, {RoboNet Collaborations},
  {Chaboyer}, {Crocker}, {Frank}, \& {Macintosh}}]{OB06109}
{Gaudi}, B.~S., {Bennett}, D.~P., {Udalski}, A., {et~al.} 2008, Science, 319,
  927

\bibitem[{{Gould}(1992)}]{Gould1992}
{Gould}, A. 1992, \apj, 392, 442

\bibitem[{{Gould}(1994)}]{1994ApJ...421L..75G}
---. 1994, \apjl, 421, L75

\bibitem[{{Gould}(2000)}]{Gould2000}
---. 2000, \apj, 542, 785

\bibitem[{{Gould}(2004)}]{Gouldpies2004}
---. 2004, \apj, 606, 319

\bibitem[{{Gould} {et~al.}(2010){Gould}, {Dong}, {Gaudi}, {Udalski}, {Bond},
  {Greenhill}, {Street}, {Dominik}, {Sumi}, {Szyma{\'n}ski}, {Han}, {Allen},
  {Bolt}, {Bos}, {Christie}, {DePoy}, {Drummond}, {Eastman}, {Gal-Yam},
  {Higgins}, {Janczak}, {Kaspi}, {Koz{\l}owski}, {Lee}, {Mallia}, {Maury},
  {Maoz}, {McCormick}, {Monard}, {Moorhouse}, {Morgan}, {Natusch}, {Ofek},
  {Park}, {Pogge}, {Polishook}, {Santallo}, {Shporer}, {Spector}, {Thornley},
  {Yee}, {{$\mu$}FUN Collaboration}, {Kubiak}, {Pietrzy{\'n}ski},
  {Soszy{\'n}ski}, {Szewczyk}, {Wyrzykowski}, {Ulaczyk}, {Poleski}, {OGLE
  Collaboration}, {Abe}, {Bennett}, {Botzler}, {Douchin}, {Freeman}, {Fukui},
  {Furusawa}, {Hearnshaw}, {Hosaka}, {Itow}, {Kamiya}, {Kilmartin}, {Korpela},
  {Lin}, {Ling}, {Makita}, {Masuda}, {Matsubara}, {Miyake}, {Muraki}, {Nagaya},
  {Nishimoto}, {Ohnishi}, {Okumura}, {Perrott}, {Philpott}, {Rattenbury},
  {Saito}, {Sako}, {Sullivan}, {Sweatman}, {Tristram}, {von Seggern}, {Yock},
  {MOA Collaboration}, {Albrow}, {Batista}, {Beaulieu}, {Brillant}, {Caldwell},
  {Calitz}, {Cassan}, {Cole}, {Cook}, {Coutures}, {Dieters}, {Dominis Prester},
  {Donatowicz}, {Fouqu{\'e}}, {Hill}, {Hoffman}, {Jablonski}, {Kane}, {Kains},
  {Kubas}, {Marquette}, {Martin}, {Martioli}, {Meintjes}, {Menzies},
  {Pedretti}, {Pollard}, {Sahu}, {Vinter}, {Wambsganss}, {Watson}, {Williams},
  {Zub}, {PLANET Collaboration}, {Allan}, {Bode}, {Bramich}, {Burgdorf},
  {Clay}, {Fraser}, {Hawkins}, {Horne}, {Kerins}, {Lister}, {Mottram},
  {Saunders}, {Snodgrass}, {Steele}, {Tsapras}, {RoboNet Collaboration},
  {J{\o}rgensen}, {Anguita}, {Bozza}, {Calchi Novati}, {Harps{\o}e}, {Hinse},
  {Hundertmark}, {Kj{\ae}rgaard}, {Liebig}, {Mancini}, {Masi}, {Mathiasen},
  {Rahvar}, {Ricci}, {Scarpetta}, {Southworth}, {Surdej}, {Th{\"o}ne}, \&
  {MiNDSTEp Consortium}}]{mufun}
{Gould}, A., {Dong}, S., {Gaudi}, B.~S., {et~al.} 2010, \apj, 720, 1073

\bibitem[{{Gould} {et~al.}(2014){Gould}, {Udalski}, {Shin}, {Porritt},
  {Skowron}, {Han}, {Yee}, {Koz{\l}owski}, {Choi}, {Poleski}, {Wyrzykowski},
  {Ulaczyk}, {Pietrukowicz}, {Mr{\'o}z}, {Szyma{\'n}ski}, {Kubiak},
  {Soszy{\'n}ski}, {Pietrzy{\'n}ski}, {Gaudi}, {Christie}, {Drummond},
  {McCormick}, {Natusch}, {Ngan}, {Tan}, {Albrow}, {DePoy}, {Hwang}, {Jung},
  {Lee}, {Park}, {Pogge}, {Abe}, {Bennett}, {Bond}, {Botzler}, {Freeman},
  {Fukui}, {Fukunaga}, {Itow}, {Koshimoto}, {Larsen}, {Ling}, {Masuda},
  {Matsubara}, {Muraki}, {Namba}, {Ohnishi}, {Philpott}, {Rattenbury}, {Saito},
  {Sullivan}, {Sumi}, {Suzuki}, {Tristram}, {Tsurumi}, {Wada}, {Yamai}, {Yock},
  {Yonehara}, {Shvartzvald}, {Maoz}, {Kaspi}, \& {Friedmann}}]{OB130341}
{Gould}, A., {Udalski}, A., {Shin}, I.~G., {et~al.} 2014, Science, 345, 46

\bibitem[{{Gould} {et~al.}(2020){Gould}, {Ryu}, {Calchi Novati}, {Zang},
  {Albrow}, {Chung}, {Han}, {Hwang}, {Jung}, {Shin}, {Shvartzvald}, {Yee},
  {Cha}, {Kim}, {Kim}, {Kim}, {Lee}, {Lee}, {Lee}, {Park}, {Pogge}, {Beichman},
  {Bryden}, {Carey}, {Gaudi}, {Henderson}, {Zhu}, {Fouque}, {Penny}, {Petric},
  {Burdullis}, \& {Mao}}]{KB180029}
{Gould}, A., {Ryu}, Y.-H., {Calchi Novati}, S., {et~al.} 2020, Journal of
  Korean Astronomical Society, 53, 9

\bibitem[{{Griest} \& {Hu}(1992)}]{Griest1992}
{Griest}, K., \& {Hu}, W. 1992, \apj, 397, 362

\bibitem[{{Griest} \& {Safizadeh}(1998)}]{Griest1998}
{Griest}, K., \& {Safizadeh}, N. 1998, \apj, 500, 37

\bibitem[{{Han}(2005)}]{Han2005}
{Han}, C. 2005, \apj, 629, 1102

\bibitem[{{Han} \& {Gould}(1997)}]{HanGould1997}
{Han}, C., \& {Gould}, A. 1997, \apj, 480, 196

\bibitem[{{Han} {et~al.}(2013){Han}, {Udalski}, {Choi}, {Yee}, {Gould},
  {Christie}, {Tan}, {Szyma{\'n}ski}, {Kubiak}, {Soszy{\'n}ski},
  {Pietrzy{\'n}ski}, {Poleski}, {Ulaczyk}, {Pietrukowicz}, {Koz{\l}owski},
  {Skowron}, {Wyrzykowski}, {OGLE Collaboration}, {Almeida}, {Batista},
  {Depoy}, {Dong}, {Drummond}, {Gaudi}, {Hwang}, {Jablonski}, {Jung}, {Lee},
  {Koo}, {McCormick}, {Monard}, {Natusch}, {Ngan}, {Park}, {Pogge}, {Porritt},
  {Shin}, \& {{\ensuremath{\mu}}FUN Collaboration}}]{OB120026}
{Han}, C., {Udalski}, A., {Choi}, J.~Y., {et~al.} 2013, \apjl, 762, L28

\bibitem[{{Han} {et~al.}(2017){Han}, {Udalski}, {Gould}, {Lee}, {Shvartzvald},
  {Zang}, {Mao}, {Koz{\l}owski}, {Albrow}, {Chung}, {Hwang}, {Jung}, {Kim},
  {Kim}, {Ryu}, {Shin}, {Yee}, {Zhu}, {Cha}, {Kim}, {Kim}, {Lee}, {Park},
  {KMTNet Collaboration}, {Skowron}, {Mr{\'o}z}, {Pietrukowicz}, {Poleski},
  {Szyma{\'n}ski}, {Soszy{\'n}ski}, {Ulaczyk}, {Pawlak}, {OGLE Collaboration},
  {Beichman}, {Bryden}, {Calchi Novati}, {Gaudi}, {Henderson}, {Howell},
  {Jacklin}, {UKIRT Microlensing Team}, {Penny}, {Fouqu{\'e}}, {Wang}, \&
  {CFHT-K2C9 Microlensing Collaboration}}]{OB160613}
{Han}, C., {Udalski}, A., {Gould}, A., {et~al.} 2017, \aj, 154, 223

\bibitem[{{Han} {et~al.}(2019{\natexlab{a}}){Han}, {Bennett}, {Udalski},
  {Gould}, {Bond}, {Shvartzvald}, {Nikolaus}, {Hundertmark}, {Bozza}, {Cassan},
  {Hirao}, {Bachelet}, {Fouqu{\'e}}, {Authors}, {Albrow}, {Chung}, {Hong},
  {Hwang}, {Lee}, {Ryu}, {Shin}, {Yee}, {Jung}, {Cha}, {Kim}, {Kim}, {Kim},
  {Kim}, {Lee}, {Lee}, {Park}, {Pogge}, {The KMTNet Collaboration}, {Mr{\'o}z},
  {Szyma{\'n}ski}, {Skowron}, {Poleski}, {Soszy{\'n}ski}, {Pietrukowicz},
  {Koz{\l}owski}, {Ulaczyk}, {Rybicki}, {Iwanek}, {Wrona}, {The OGLE
  Collaboration}, {Abe}, {Barry}, {Bhattacharya}, {Donachie}, {Fukui}, {Itow},
  {Kawasaki}, {Kondo}, {Koshimoto}, {Li}, {Matsubara}, {Muraki}, {Miyazaki},
  {Nagakane}, {Ranc}, {Rattenbury}, {Suematsu}, {Sullivan}, {Sumi}, {Suzuki},
  {Tristram}, {Yonehara}, {The MOA Collaboration}, {Mao}, {Wang}, {Zang},
  {Zhu}, {Penny}, {The CFHT Collaboration}, {Beichman}, {Bryden}, {Calchi
  Novati}, {Gaudi}, {Henderson}, {Jacklin}, {Stassun}, \& {The UKIRT
  Microlensing Team}}]{OB181011}
{Han}, C., {Bennett}, D.~P., {Udalski}, A., {et~al.} 2019{\natexlab{a}}, \aj,
  158, 114

\bibitem[{{Han} {et~al.}(2019{\natexlab{b}}){Han}, {Yee}, {Udalski}, {Bond},
  {Bozza}, {Cassan}, {Hirao}, {Dong}, {Kollmeier}, {Morrell}, {Boutsia},
  {authors}, {Albrow}, {Chung}, {Gould}, {Hwang}, {Lee}, {Ryu}, {Shin},
  {Shvartzvald}, {Jung}, {Kim}, {Kim}, {Cha}, {Kim}, {Kim}, {Hong}, {Kim},
  {Lee}, {Lee}, {Park}, {Pogge}, {Zang}, {The KMTNet Collaboration},
  {Mr{\'o}z}, {Szyma{\'n}ski}, {Skowron}, {Poleski}, {Soszy{\'n}ski},
  {Pietrukowicz}, {Koz{\l}owski}, {Ulaczyk}, {Rybicki}, {Iwanek}, {Wrona}, {The
  OGLE Collaboration}, {Abe}, {Barry}, {Bennett}, {Bhattacharya}, {Donachie},
  {Fukui}, {Itow}, {Kawasaki}, {Kondo}, {Koshimoto}, {Li}, {Matsubara},
  {Muraki}, {Miyazaki}, {Nagakane}, {Ranc}, {Rattenbury}, {Suematsu},
  {Sullivan}, {Sumi}, {Suzuki}, {Tristram}, {Yonehara}, \& {The MOA
  Collaboration}}]{OB180740}
{Han}, C., {Yee}, J.~C., {Udalski}, A., {et~al.} 2019{\natexlab{b}}, \aj, 158,
  102

\bibitem[{{Han} {et~al.}(2020{\natexlab{a}}){Han}, {Shin}, {Jung}, {Kim},
  {Yee}, {Albrow}, {Chung}, {Gould}, {Hwang}, {Lee}, {Ryu}, {Shvartzvald},
  {Zang}, {Cha}, {Kim}, {Kim}, {Kim}, {Lee}, {Lee}, {Park}, \&
  {Pogge}}]{KB180748}
{Han}, C., {Shin}, I.-G., {Jung}, Y.~K., {et~al.} 2020{\natexlab{a}}, \aap,
  641, A105

\bibitem[{{Han} {et~al.}(2020{\natexlab{b}}){Han}, {Kim}, {Jung}, {Gould},
  {Bond}, {Albrow}, {Chung}, {Hwang}, {Lee}, {Ryu}, {Shin}, {Shvartzvald},
  {Yee}, {Zang}, {Cha}, {Kim}, {Kim}, {Kim}, {Lee}, {Lee}, {Park}, {Pogge},
  {Kim}, {KMTNet Collaboration}, {Abe}, {Barry}, {Bennett}, {Bhattacharya},
  {Donachie}, {Fujii}, {Fukui}, {Itow}, {Hirao}, {Kirikawa}, {Kondo},
  {Koshimoto}, {Li}, {Matsubara}, {Muraki}, {Miyazaki}, {Nagakane}, {Ranc},
  {Rattenbury}, {Satoh}, {Shoji}, {Suematsu}, {Sumi}, {Suzuki}, {Tanaka},
  {Tristram}, {Yamawaki}, {Yonehara}, \& {MOA Collaboration}}]{KB191953}
{Han}, C., {Kim}, D., {Jung}, Y.~K., {et~al.} 2020{\natexlab{b}}, \aj, 160, 17

\bibitem[{{Han} {et~al.}(2021){Han}, {Lee}, {Ryu}, {Kim}, {Albrow}, {Chung},
  {Gould}, {Hwang}, {Jung}, {Kim}, {Shin}, {Shvartzvald}, {Yee}, {Zang}, {Cha},
  {Kim}, {Kim}, {Lee}, {Lee}, {Park}, \& {Pogge}}]{KB190797}
{Han}, C., {Lee}, C.-U., {Ryu}, Y.-H., {et~al.} 2021, arXiv e-prints,
  arXiv:2102.01806

\bibitem[{{Hwang} {et~al.}(2018){Hwang}, {Udalski}, {Shvartzvald}, {Ryu},
  {Albrow}, {Chung}, {Gould}, {Han}, {Jung}, {}, {Yee}, {Zhu}, {Cha}, {Kim},
  {Kim}, {Kim}, {Lee}, {Lee}, {Lee}, {Park}, {Pogge}, {KMTNet Collaboration},
  {Skowron}, {Mr{\'o}z}, {Poleski}, {Koz{\l}owski}, {Soszy{\'n}ski},
  {Pietrukowicz}, {Szyma{\'n}ski}, {Ulaczyk}, {Pawlak}, {OGLE Collaboration},
  {Bryden}, {Beichman}, {Calchi Novati}, {Gaudi}, {Henderson}, {Jacklin},
  {Penny}, \& {UKIRT Microlensing Team}}]{OB170173}
{Hwang}, K.-H., {Udalski}, A., {Shvartzvald}, Y., {et~al.} 2018, \aj, 155, 20

\bibitem[{{Jiang} {et~al.}(2004){Jiang}, {DePoy}, {Gal-Yam}, {Gaudi}, {Gould},
  {Han}, {Lipkin}, {Maoz}, {Ofek}, {Park}, {Pogge}, {MuFun Collaboration},
  {Udalski}, {Kubiak}, {Szyma{\'n}ski}, {Szewczyk}, {{\.Z}ebru{\'n}},
  {Wyrzykowski}, {Soszy{\'n}ski}, {Pietrzy{\'n}ski}, {OGLE Collaboration},
  {Albrow}, {Beaulieu}, {Caldwell}, {Cassan}, {Coutures}, {Dominik},
  {Donatowicz}, {Fouqu{\'e}}, {Greenhill}, {Hill}, {Horne}, {J{\o}rgensen},
  {J{\o}rgensen}, {Kane}, {Kubas}, {Martin}, {Menzies}, {Pollard}, {Sahu},
  {Wambsganss}, {Watson}, {Williams}, \& {PLANET Collaboration}}]{Jiang2004}
{Jiang}, G., {DePoy}, D.~L., {Gal-Yam}, A., {et~al.} 2004, \apj, 617, 1307

\bibitem[{{Jung} {et~al.}(2019){Jung}, {Gould}, {Zang}, {Hwang}, {Ryu}, {Han},
  {Yee}, {Albrow}, {Chung}, {Shin}, {Shvartzvald}, {Zhu}, {Cha}, {Kim}, {Kim},
  {Kim}, {Lee}, {Lee}, {Lee}, {Park}, {Pogge}, {The KMTNet Collaboration},
  {Penny}, {Mao}, {Fouqu{\'e}}, {Wang}, \& {The CFHT Collaboration}}]{KB170165}
{Jung}, Y.~K., {Gould}, A., {Zang}, W., {et~al.} 2019, \aj, 157, 72

\bibitem[{{Jung} {et~al.}(2020){Jung}, {Udalski}, {Zang}, {Bond}, {Yee}, {Han},
  {Albrow}, {Chung}, {Gould}, {Hwang}, {Ryu}, {Shin}, {Shvartzvald}, {Cha},
  {Kim}, {Kim}, {Kim}, {Lee}, {Lee}, {Lee}, {Park}, {Pogge}, {KMTNet
  Collaboration}, {Mr{\'o}z}, {Szyma{\'n}ski}, {Skowron}, {Poleski},
  {Soszy{\'n}ski}, {Pietrukowicz}, {Koz{\l}owski}, {Ulaczyk}, {Rybicki},
  {Iwanek}, {Wrona}, {OGLE Collaboration}, {Abe}, {Barry}, {Bennett},
  {Bhattacharya}, {Donachie}, {Fujii}, {Fukui}, {Hirao}, {Itow}, {Kamei},
  {Kondo}, {Koshimoto}, {Li}, {Matsubara}, {Miyazaki}, {Muraki}, {Nagakane},
  {Ranc}, {Rattenbury}, {Satoh}, {Shoji}, {Suematsu}, {Sullivan}, {Sumi},
  {Suzuki}, {Tristram}, {Yamakawa}, {Yamamwaki}, {Yonehara}, \& {MOA
  Collaboration}}]{KB190842}
{Jung}, Y.~K., {Udalski}, A., {Zang}, W., {et~al.} 2020, \aj, 160, 255

\bibitem[{{Kim} {et~al.}(2018){Kim}, {Hwang}, {Shvartzvald}, {Yee}, {Albrow},
  {Cha}, {Chung}, {Gould}, {Han}, {Jung}, {Kim}, {Kim}, {Lee}, {Lee}, {Lee},
  {Park}, {Pogge}, {Ryu}, {Shin}, \& {Zang}}]{KMTAF}
{Kim}, H.-W., {Hwang}, K.-H., {Shvartzvald}, Y., {et~al.} 2018, arXiv e-prints,
  arXiv:1806.07545

\bibitem[{{Kim} {et~al.}(2016){Kim}, {Lee}, {Park}, {Kim}, {Cha}, {Lee}, {Han},
  {Chun}, \& {Yuk}}]{KMT2016}
{Kim}, S.-L., {Lee}, C.-U., {Park}, B.-G., {et~al.} 2016, Journal of Korean
  Astronomical Society, 49, 37

\bibitem[{{Koz{\l}owski} {et~al.}(2007){Koz{\l}owski}, {Wo{\'z}niak}, {Mao}, \&
  {Wood}}]{Kozlowski2007}
{Koz{\l}owski}, S., {Wo{\'z}niak}, P.~R., {Mao}, S., \& {Wood}, A. 2007, \apj,
  671, 420

\bibitem[{Madsen \& Zhu(2019)}]{OB120026_Zhu}
Madsen, S., \& Zhu, W. 2019, \aj, 878, L29

\bibitem[{{Nataf} {et~al.}(2013){Nataf}, {Gould}, {Fouqu{\'e}}, {Gonzalez},
  {Johnson}, {Skowron}, {}, {Szyma{\'n}ski}, {Kubiak}, {Pietrzy{\'n}ski},
  {Soszy{\'n}ski}, {Ulaczyk}, {Wyrzykowski}, \& {Poleski}}]{Nataf2013}
{Nataf}, D.~M., {Gould}, A., {Fouqu{\'e}}, P., {et~al.} 2013, \apj, 769, 88

\bibitem[{{Nemiroff} \& {Wickramasinghe}(1994)}]{Nemiroff1994}
{Nemiroff}, R.~J., \& {Wickramasinghe}, W.~A.~D.~T. 1994, \apjl, 424, L21

\bibitem[{{Paczy{\'n}ski}(1986)}]{Paczynski1986}
{Paczy{\'n}ski}, B. 1986, \apj, 304, 1

\bibitem[{{Park} {et~al.}(2004){Park}, {DePoy}, {Gaudi}, {Gould}, {Han},
  {Pogge}, {muFun Collaboration}, {Abe}, {Bennett}, {Bond}, {Eguchi}, {Furuta},
  {Hearnshaw}, {Kamiya}, {Kilmartin}, {Kurata}, {Masuda}, {Matsubara},
  {Muraki}, {Noda}, {Okajima}, {Rattenbury}, {Sako}, {Sekiguchi}, {Sullivan},
  {Sumi}, {Tristram}, {Yanagisawa}, {Yock}, \& {MOA Collaboration}}]{MB03037}
{Park}, B.~G., {DePoy}, D.~L., {Gaudi}, B.~S., {et~al.} 2004, \apj, 609, 166

\bibitem[{{Poindexter} {et~al.}(2005){Poindexter}, {Afonso}, {Bennett},
  {Glicenstein}, {Gould}, {Szyma{\'n}ski}, \& {Udalski}}]{Poindexter2005}
{Poindexter}, S., {Afonso}, C., {Bennett}, D.~P., {et~al.} 2005, \apj, 633, 914

\bibitem[{{Reid} {et~al.}(2014){Reid}, {Menten}, {Brunthaler}, {Zheng}, {Dame},
  {Xu}, {Wu}, {Zhang}, {Sanna}, {Sato}, {Hachisuka}, {Choi}, {Immer},
  {Moscadelli}, {Rygl}, \& {Bartkiewicz}}]{Reid2014}
{Reid}, M.~J., {Menten}, K.~M., {Brunthaler}, A., {et~al.} 2014, \apj, 783, 130

\bibitem[{{Rowe} {et~al.}(2015){Rowe}, {Jarvis}, {Mandelbaum}, {Bernstein},
  {Bosch}, {Simet}, {Meyers}, {Kacprzak}, {Nakajima}, {Zuntz}, {Miyatake},
  {Dietrich}, {Armstrong}, {Melchior}, \& {Gill}}]{GalSim}
{Rowe}, B.~T.~P., {Jarvis}, M., {Mandelbaum}, R., {et~al.} 2015, Astronomy and
  Computing, 10, 121

\bibitem[{{Ryu} {et~al.}(2020{\natexlab{a}}){Ryu}, {Navarro}, {Gould},
  {Albrow}, {Chung}, {Han}, {Hwang}, {Jung}, {Shin}, {Shvartzvald}, {Yee},
  {Zang}, {Cha}, {Kim}, {Kim}, {Kim}, {Lee}, {Lee}, {Lee}, {Park}, {Pogge},
  {Minniti}, {Saito}, {Alonso-Garc{\'\i}a}, \& {Penny}}]{KB181292}
{Ryu}, Y.-H., {Navarro}, M.~G., {Gould}, A., {et~al.} 2020{\natexlab{a}}, \aj,
  159, 58

\bibitem[{{Ryu} {et~al.}(2020{\natexlab{b}}){Ryu}, {Udalski}, {Yee}, {Penny},
  {Zang}, {Albrow}, {Chung}, {Gould}, {Han}, {Hwang}, {Jung}, {Shin},
  {Shvartzvald}, {Cha}, {Kim}, {Kim}, {Kim}, {Lee}, {Lee}, {Lee}, {Park},
  {Pogge}, {KMTNet Collaboration}, {Mr{\'o}z}, {Szyma{\'n}ski}, {Skowron},
  {Poleski}, {Soszy{\'n}ski}, {Pietrukowicz}, {Koz{\l}owski}, {Ulaczyk},
  {Rybicki}, {Iwanek}, {Wrona}, {OGLE Collaboration}, {Mao}, {Fouque}, {Zhu},
  {Wang}, \& {CFHT Microlensing Collaboration}}]{OB180532}
{Ryu}, Y.-H., {Udalski}, A., {Yee}, J.~C., {et~al.} 2020{\natexlab{b}}, \aj,
  160, 183

\bibitem[{{Schechter} {et~al.}(1993){Schechter}, {Mateo}, \& {Saha}}]{dophot}
{Schechter}, P.~L., {Mateo}, M., \& {Saha}, A. 1993, \pasp, 105, 1342

\bibitem[{{Skowron} {et~al.}(2011){Skowron}, {Udalski}, {Gould}, {Dong},
  {Monard}, {Han}, {Nelson}, {McCormick}, {Moorhouse}, {Thornley}, {Maury},
  {Bramich}, {Greenhill}, {Koz{\l}owski}, {Bond}, {Poleski}, {Wyrzykowski},
  {Ulaczyk}, {Kubiak}, {Szyma{\'n}ski}, {Pietrzy{\'n}ski}, {Soszy{\'n}ski},
  {OGLE Collaboration}, {Gaudi}, {Yee}, {Hung}, {Pogge}, {DePoy}, {Lee},
  {Park}, {Allen}, {Mallia}, {Drummond}, {Bolt}, {{$\mu$}FUN Collaboration},
  {Allan}, {Browne}, {Clay}, {Dominik}, {Fraser}, {Horne}, {Kains}, {Mottram},
  {Snodgrass}, {Steele}, {Street}, {Tsapras}, {RoboNet Collaboration}, {Abe},
  {Bennett}, {Botzler}, {Douchin}, {Freeman}, {Fukui}, {Furusawa}, {Hayashi},
  {Hearnshaw}, {Hosaka}, {Itow}, {Kamiya}, {Kilmartin}, {Korpela}, {Lin},
  {Ling}, {Makita}, {Masuda}, {Matsubara}, {Muraki}, {Nagayama}, {Miyake},
  {Nishimoto}, {Ohnishi}, {Perrott}, {Rattenbury}, {Saito}, {Skuljan},
  {Sullivan}, {Sumi}, {Suzuki}, {Sweatman}, {Tristram}, {Wada}, {Yock}, {MOA
  Collaboration}, {Beaulieu}, {Fouqu{\'e}}, {Albrow}, {Batista}, {Brillant},
  {Caldwell}, {Cassan}, {Cole}, {Cook}, {Coutures}, {Dieters}, {Dominis
  Prester}, {Donatowicz}, {Kane}, {Kubas}, {Marquette}, {Martin}, {Menzies},
  {Sahu}, {Wambsganss}, {Williams}, {Zub}, \& {PLANET Collaboration}}]{OB09020}
{Skowron}, J., {Udalski}, A., {Gould}, A., {et~al.} 2011, \apj, 738, 87

\bibitem[{{Sumi} {et~al.}(2016){Sumi}, {Udalski}, {Bennett}, {Gould},
  {Poleski}, {Bond}, {Skowron}, {Rattenbury}, {Pogge}, {Bensby}, {Beaulieu},
  {Marquette}, {Batista}, {Brillant}, {Abe}, {Asakura}, {Bhattacharya},
  {Donachie}, {Freeman}, {Fukui}, {Hirao}, {Itow}, {Koshimoto}, {Li}, {Ling},
  {Masuda}, {Matsubara}, {Muraki}, {Nagakane}, {Ohnishi}, {Oyokawa}, {Saito},
  {Sharan}, {Sullivan}, {Suzuki}, {Tristram}, {Yonehara}, {MOA Collaboration},
  {Szyma{\'n}ski}, {Ulaczyk}, {Koz{\l}owski}, {Wyrzykowski}, {Kubiak},
  {Pietrukowicz}, {Pietrzy{\'n}ski}, {Soszy{\'n}ski}, {OGLE Collaboration},
  {Han}, {Jung}, {}, \& {Lee}}]{MOA2016}
{Sumi}, T., {Udalski}, A., {Bennett}, D.~P., {et~al.} 2016, \apj, 825, 112

\bibitem[{{Suzuki} {et~al.}(2016){Suzuki}, {Bennett}, {Sumi}, {Bond}, {Rogers},
  {Abe}, {Asakura}, {Bhattacharya}, {Donachie}, {Freeman}, {Fukui}, {Hirao},
  {Itow}, {Koshimoto}, {Li}, {Ling}, {Masuda}, {Matsubara}, {Muraki},
  {Nagakane}, {Onishi}, {Oyokawa}, {Rattenbury}, {Saito}, {Sharan}, {Shibai},
  {Sullivan}, {Tristram}, {Yonehara}, \& {MOA Collaboration}}]{Suzuki2016}
{Suzuki}, D., {Bennett}, D.~P., {Sumi}, T., {et~al.} 2016, \apj, 833, 145

\bibitem[{{Suzuki} {et~al.}(2018){Suzuki}, {Bennett}, {Udalski}, {Bond},
  {Sumi}, {Han}, {Kim}, {Abe}, {Asakura}, {Barry}, {Bhattacharya}, {Donachie},
  {Freeman}, {Fukui}, {Hirao}, {Itow}, {Koshimoto}, {Li}, {Ling}, {Masuda},
  {Matsubara}, {Muraki}, {Nagakane}, {Onishi}, {Oyokawa}, {Ranc}, {Rattenbury},
  {Saito}, {Sharan}, {Sullivan}, {Tristram}, {Yonehara}, {MOA Collaboration},
  {Poleski}, {Mr{\'o}z}, {Skowron}, {Szyma{\'n}ski}, {Soszy{\'n}ski},
  {Koz{\l}owski}, {Pietrukowicz}, {Wyrzykowski}, {Ulaczyk}, \& {OGLE
  Collaboration}}]{OB141722}
{Suzuki}, D., {Bennett}, D.~P., {Udalski}, A., {et~al.} 2018, \aj, 155, 263

\bibitem[{{Szyma{\'n}ski} {et~al.}(2011){Szyma{\'n}ski}, {Udalski},
  {Soszy{\'n}ski}, {Kubiak}, {Pietrzy{\'n}ski}, {Poleski}, {Wyrzykowski}, \&
  {Ulaczyk}}]{OGLEIII}
{Szyma{\'n}ski}, M.~K., {Udalski}, A., {Soszy{\'n}ski}, I., {et~al.} 2011,
  \actaa, 61, 83

\bibitem[{{Tomaney} \& {Crotts}(1996)}]{Tomaney1996}
{Tomaney}, A.~B., \& {Crotts}, A. P.~S. 1996, \aj, 112, 2872

\bibitem[{{Udalski} {et~al.}(2015){Udalski}, {Szyma{\'n}ski}, \&
  {Szyma{\'n}ski}}]{OGLEIV}
{Udalski}, A., {Szyma{\'n}ski}, M.~K., \& {Szyma{\'n}ski}, G. 2015, \actaa, 65,
  1

\bibitem[{{Udalski} {et~al.}(2018){Udalski}, {Ryu}, {Sajadian}, {Gould},
  {Mr{\'o}z}, {Poleski}, {Szyma{\'n}ski}, {Skowron}, {Soszy{\'n}ski},
  {Koz{\l}owski}, {Pietrukowicz}, {Ulaczyk}, {Pawlak}, {Rybicki}, {Iwanek},
  {Albrow}, {Chung}, {Han}, {Hwang}, {Jung}, {}, {Shvartzvald}, {Yee}, {Zang},
  {Zhu}, {Cha}, {Kim}, {Kim}, {Kim}, {Lee}, {Lee}, {Lee}, {Park}, {Pogge},
  {Bozza}, {Dominik}, {Helling}, {Hundertmark}, {J{\o}rgensen},
  {Longa-Pe{\~n}a}, {Lowry}, {Burgdorf}, {Campbell-White}, {Ciceri}, {Evans},
  {Figuera Jaimes}, {Fujii}, {Haikala}, {Henning}, {Hinse}, {Mancini},
  {Peixinho}, {Rahvar}, {Rabus}, {Skottfelt}, {Snodgrass}, {Southworth}, \&
  {von Essen}}]{OB171434}
{Udalski}, A., {Ryu}, Y.-H., {Sajadian}, S., {et~al.} 2018, \actaa, 68, 1

\bibitem[{{Wang} {et~al.}(2018){Wang}, {Calchi Novati}, {Udalski}, {Gould},
  {Mao}, {Zang}, {Beichman}, {Bryden}, {Carey}, {Gaudi}, {Henderson},
  {Shvartzvald}, {Yee}, {Spitzer Team}, {Mr{\'o}z}, {Poleski}, {Skowron},
  {Szyma{\'n}ski}, {Soszy{\'n}ski}, {Koz{\l}owski}, {Pietrukowicz}, {Ulaczyk},
  {Pawlak}, {OGLE Collaboration}, {Albrow}, {Chung}, {Han}, {Hwang}, {Jung},
  {Ryu}, {Shin}, {Zhu}, {Cha}, {Kim}, {Kim}, {Kim}, {Lee}, {Lee}, {Lee},
  {Park}, {Pogge}, \& {KMTNet Collaboration}}]{OB171130}
{Wang}, T., {Calchi Novati}, S., {Udalski}, A., {et~al.} 2018, \apj, 860, 25

\bibitem[{{Witt} \& {Mao}(1994)}]{Shude1994}
{Witt}, H.~J., \& {Mao}, S. 1994, \apj, 430, 505

\bibitem[{{Yee} {et~al.}(2009){Yee}, {Udalski}, {Sumi}, {Dong}, {Koz{\l}owski},
  {Bird}, {Cole}, {Higgins}, {McCormick}, {Monard}, {Polishook}, {Shporer},
  {Spector}, {Szyma{\'n}ski}, {Kubiak}, {Pietrzy{\'n}ski}, {Soszy{\'n}ski},
  {Szewczyk}, {Ulaczyk}, {Wyrzykowski}, {Poleski}, {OGLE Collaboration},
  {Allen}, {Bos}, {Christie}, {DePoy}, {Eastman}, {Gaudi}, {Gould}, {Han},
  {Kaspi}, {Lee}, {Mallia}, {Maury}, {Maoz}, {Natusch}, {Park}, {Pogge},
  {Santallo}, {{$\mu$}FUN Collaboration}, {Abe}, {Bond}, {Fukui}, {Furusawa},
  {Hearnshaw}, {Hosaka}, {Itow}, {Kamiya}, {Korpela}, {Kilmartin}, {Lin},
  {Ling}, {Makita}, {Masuda}, {Matsubara}, {Miyake}, {Muraki}, {Nagaya},
  {Nishimoto}, {Ohnishi}, {Perrott}, {Rattenbury}, {Sako}, {Saito}, {Skuljan},
  {Sullivan}, {Sweatman}, {Tristram}, {Yock}, {MOA Collaboration}, {Albrow},
  {Batista}, {Fouqu{\'e}}, {Beaulieu}, {Bennett}, {Cassan}, {Comparat},
  {Coutures}, {Dieters}, {Greenhill}, {Horne}, {Kains}, {Kubas}, {Martin},
  {Menzies}, {Wambsganss}, {Williams}, {Zub}, \& {PLANET
  Collaboration}}]{OB08279}
{Yee}, J.~C., {Udalski}, A., {Sumi}, T., {et~al.} 2009, \apj, 703, 2082

\bibitem[{{Yee} {et~al.}(2012){Yee}, {Shvartzvald}, {Gal-Yam}, {Bond},
  {Udalski}, {Koz{\l}owski}, {Han}, {Gould}, {Skowron}, {Suzuki}, {Abe},
  {Bennett}, {Botzler}, {Chote}, {Freeman}, {Fukui}, {Furusawa}, {Itow},
  {Kobara}, {Ling}, {Masuda}, {Matsubara}, {Miyake}, {Muraki}, {Ohmori},
  {Ohnishi}, {Rattenbury}, {Saito}, {Sullivan}, {Sumi}, {Suzuki}, {Sweatman},
  {Takino}, {Tristram}, {Wada}, {MOA Collaboration}, {Szyma{\'n}ski}, {Kubiak},
  {Pietrzy{\'n}ski}, {Soszy{\'n}ski}, {Poleski}, {Ulaczyk}, {Wyrzykowski},
  {Pietrukowicz}, {OGLE Collaboration}, {Allen}, {Almeida}, {Batista}, {Bos},
  {Christie}, {DePoy}, {Dong}, {Drummond}, {Finkelman}, {Gaudi}, {Gorbikov},
  {Henderson}, {Higgins}, {Jablonski}, {Kaspi}, {Manulis}, {Maoz}, {McCormick},
  {McGregor}, {Monard}, {Moorhouse}, {Mu{\~n}oz}, {Natusch}, {Ngan}, {Ofek},
  {Pogge}, {Santallo}, {Tan}, {Thornley}, {Shin}, {Choi}, {Park}, {Lee}, {Koo},
  \& {{\ensuremath{\mu}}FUN Collaboration}}]{MB11293}
{Yee}, J.~C., {Shvartzvald}, Y., {Gal-Yam}, A., {et~al.} 2012, \apj, 755, 102

\bibitem[{{Yee} {et~al.}(2021){Yee}, {Zang}, {Udalski}, {Ryu}, {Green},
  {Hennerley}, {Marmont}, {Sumi}, {Mao}, {Gromadzki}, {Mr{\'o}z}, {Skowron},
  {Poleski}, {Szyma{\'n}ski}, {Soszy{\'n}ski}, {Pietrukowicz}, {Koz{\l}owski},
  {Ulaczyk}, {Rybicki}, {Iwanek}, {Wrona}, {Albrow}, {Chung}, {Gould}, {Han},
  {Hwang}, {Jung}, {Kim}, {Shin}, {Shvartzvald}, {Cha}, {Kim}, {Kim}, {Lee},
  {Lee}, {Lee}, {Park}, {Pogge}, {Bachelet}, {Christie}, {Hundertmark}, {Maoz},
  {McCormick}, {Natusch}, {Penny}, {Street}, {Tsapras}, {Beichman}, {Bryden},
  {Calchi Novati}, {Carey}, {Gaudi}, {Henderson}, {Johnson}, {Zhu}, {Bond},
  {Abe}, {Barry}, {Bennett}, {Bhattacharya}, {Donachie}, {Fujii}, {Fukui},
  {Hirao}, {Ishitani Silva}, {Itow}, {Kirikawa}, {Kondo}, {Koshimoto}, {Li},
  {Matsubara}, {Muraki}, {Miyazaki}, {Olmschenk}, {Ranc}, {Rattenbury},
  {Satoh}, {Shoji}, {Suzuki}, {Tanaka}, {Tristram}, {Yamawaki}, \&
  {Yonehara}}]{OB190960}
{Yee}, J.~C., {Zang}, W., {Udalski}, A., {et~al.} 2021, arXiv e-prints,
  arXiv:2101.04696

\bibitem[{{Yoo} {et~al.}(2004){Yoo}, {DePoy}, {Gal-Yam}, {Gaudi}, {Gould},
  {Han}, {Lipkin}, {Maoz}, {Ofek}, {Park}, {Pogge}, {Mu-Fun Collaboration},
  {Udalski}, {Soszy{\'n}ski}, {Wyrzykowski}, {Kubiak}, {Szyma{\'n}ski},
  {Pietrzy{\'n}ski}, {Szewczyk}, {{\.Z}ebru{\'n}}, \& {OGLE
  Collaboration}}]{Yoo2004}
{Yoo}, J., {DePoy}, D.~L., {Gal-Yam}, A., {et~al.} 2004, \apj, 603, 139

\bibitem[{{Zang} {et~al.}(2018){Zang}, {Penny}, {Zhu}, {Mao}, {Fouqu{\'e}},
  {Udalski}, {Hwang}, {Wang}, {Huang}, {Boyajian}, \& {Barentsen}}]{CFHT}
{Zang}, W., {Penny}, M.~T., {Zhu}, W., {et~al.} 2018, \pasp, 130, 104401

\bibitem[{{Zhu} {et~al.}(2014{\natexlab{a}}){Zhu}, {Gould}, {Penny}, {Mao}, \&
  {Gendron}}]{Zhu2014ApJb}
{Zhu}, W., {Gould}, A., {Penny}, M., {Mao}, S., \& {Gendron}, R.
  2014{\natexlab{a}}, \apj, 794, 53

\bibitem[{{Zhu} {et~al.}(2014{\natexlab{b}}){Zhu}, {Penny}, {Mao}, {Gould}, \&
  {Gendron}}]{Zhu2014ApJ}
{Zhu}, W., {Penny}, M., {Mao}, S., {Gould}, A., \& {Gendron}, R.
  2014{\natexlab{b}}, \apj, 788, 73

\bibitem[{{Zhu} {et~al.}(2017){Zhu}, {Udalski}, {Calchi Novati}, {Chung},
  {Jung}, {Ryu}, {}, {Gould}, {Lee}, {Albrow}, {Yee}, {Han}, {Hwang}, {Cha},
  {Kim}, {Kim}, {Kim}, {Kim}, {Lee}, {Park}, {Pogge}, {KMTNet Collaboration},
  {Poleski}, {Mr{\'o}z}, {Pietrukowicz}, {Skowron}, {Szyma{\'n}ski},
  {Koz{\l}owski}, {Ulaczyk}, {Pawlak}, {OGLE Collaboration}, {Beichman},
  {Bryden}, {Carey}, {Fausnaugh}, {Gaudi}, {Henderson}, {Shvartzvald},
  {Wibking}, \& {Spitzer Team}}]{Zhu2017spitzer}
{Zhu}, W., {Udalski}, A., {Calchi Novati}, S., {et~al.} 2017, \aj, 154, 210

\end{thebibliography}
